\documentclass[a4paper,fleqn,usenatbib,useAMS,onecolumn]{mnras}

\usepackage{graphicx}	
\usepackage{amsmath}	
\usepackage{amssymb}	
\usepackage{multicol}        
\usepackage{bm}		
\usepackage{pdflscape}	
\usepackage{dcolumn}   
\usepackage{grffile}
\usepackage{color}
\usepackage{grffile}

\newcommand\Eqn[1]     {Eq.\,(\ref{#1})}
\newcommand\Eqns[2]    {Eqs\,(\ref{#1}) and~(\ref{#2})}

\newcommand\Fig[1]     {Figure\,{\ref{#1}}}

\newcommand\nn         {\nonumber}

\def\mnras{{Mon.~ Not.~ R.~ Astron.~ Soc.~}}

\def\prd{{Phys.~ Rev.~ D.~}}

\def\apj{{Astrophys.~ J.~}}
\def\apjs{{Astrophys.~ J.~ Suppl.~}}
\def\apjl{{Astrophys.~ J.~ Lett.~}}

\def\nat{{Nature (London)~}}

\def\mnras{{MNRAS}}

\def\prd{{PRD}}

\def\apj{{ApJ}}
\def\apjs{{ApJS}}
\def\apjl{{ApJL}}
\def\aap{{A\&A}}
\def\nat{{Nature}}

\def\araa{Anual Reviews of Astronomy \& Astrophys.}

\newcommand{\be}{\begin{equation}}
\newcommand{\ee}{\end{equation}}
\newcommand{\ba}{\begin{eqnarray}}
\newcommand{\ea}{\end{eqnarray}}

\def\pp1{{\prime}}
\def\pp2{{\prime\prime}}

\def\2D{{\rm 2D}}

\def\Vmu{{V_{\mu}}}

\def\Vu{{V_{\mu}}}

\def\SN{{\mathcal S}/{\mathcal N}}

\def\c{{\rm c}}

\def\bx{{\bf x}}
\def\br{{\bf r}}

\def\bk{{\bf k}}
\def\bq{{\bf q}}

\def\Mlim{{M_{\rm Lim}}}

\def\1Loop{{\rm 1Loop}}

\def\Msol{h^{-1}M_{\odot}}

\def\kMpc{\, h \, {\rm Mpc}^{-1}}

\def\dx{{\rm d}{\bf x}}
\def\dk{{\rm d}{\bf k}}
\def\dq{{\rm d}{\bf q}}

\def\dM{{\rm d}M}

\def\nbar{\bar{n}}

\def\nbarh{\bar{n}_{\rm h}}

\def\la{\mathrel{\mathpalette\fun <}}
\def\ga{\mathrel{\mathpalette\fun >}}
\def\fun#1#2{\lower3.6pt\vbox{\baselineskip0pt\lineskip.9pt
        \ialign{$\mathsurround=0pt#1\hfill##\hfil$\crcr#2\crcr\sim\crcr}}}

\def\Fc{{\mathcal F}_{\rm c}}
\def\Fcf{\widetilde{\mathcal F}_{\rm c}}
\def\Gf{\widetilde{\mathcal G}}
\def\G{{\mathcal G}}
\def\Gb{\overline{\G}}
\def\Qf{\widetilde{\mathcal Q}}
\def\Q{{\mathcal Q}}

\def\ex{{\rm e}}
\def\Mlim{M_{\rm Lim}}
\def\Mlimx{M_{\rm Lim}(\chi)}
\def\bbar{\overline{b}}
\def\bbarsq{\overline{b^2}}
\def\bbarsqx{\overline{b^2}(\chi)}
\def\bbarM{\overline{b_M}}

\def\rhobhx{\bar{\rho}_{\rm h}(\chi)}
\def\nbarhx{\bar{n}_{\rm h}(\chi)}
\def\Omu{\Omega_{\mu}}

\def\SN{{\mathcal S}/{\mathcal N}}
\def\SNpar{\left(\frac{\mathcal S}{\mathcal N}\right)}
\def\NS{{\mathcal N}/{\mathcal S}}

\title[Toward optimal cluster power spectrum analysis]
{Toward optimal cluster power spectrum analysis}

\author[{\it Smith \& Marian}]
{Robert E.~Smith$^{1,2}$\thanks{res@mpa-garching.mpg.de,r.e.smith@sussex.ac.uk}
and Laura Marian$^{1,3}$\thanks{l.marian@sussex.ac.uk}
\\
$^1$ Max-Planck-Institut f\"ur Astrophysik,Karl-Schwarzschild-Str.1, Postfach 1523, 
85740 Garching, Germany\\
$^2$ Department of Physics and Astronomy, University of Sussex, Brighton BN1 9QH, UK\\
$^3$ Argelander-Institute for Astronomy, Auf dem H\"ugel 71, D-53121 Bonn, Germany 
}

\begin{document}

\label{firstpage}
\pagerange{\pageref{firstpage}--\pageref{lastpage}}


\maketitle

\begin{abstract}
The power spectrum of galaxy clusters is an important probe of the
cosmological model. In this paper we develop a formalism to compute
the optimal weights for the estimation of the matter power spectrum
from cluster power spectrum measurements. We find a closed-form
analytic expression for the optimal weights, which takes into account:
the cluster mass, finite survey volume effects, survey masking, and a
flux limit. The optimal weights are: $w(M,\chi) \propto
b(M,\chi)/[1+\nbarh(\chi) \bbarsqx \overline{P}(k)]$, where
$b(M,\chi)$ is the bias of clusters of mass $M$ at radial position
$\chi(z)$, $\nbarh(\chi)$ and $\bbarsqx$ are the expected space
density and bias squared of all clusters, and $\overline{P}(k)$ is the
matter power spectrum at wavenumber $k$. This result is analogous to
that of \citet{Percivaletal2004a}. We compare our optimal weighting
scheme with mass weighting and also with the original power spectrum
scheme of \citet{Feldmanetal1994}.  We show that our optimal weighting
scheme outperforms these approaches for both volume- and flux-limited
cluster surveys. Finally, we present a new expression for the Fisher
information matrix for cluster power spectrum analysis. Our expression
shows that for an optimally weighted cluster survey the cosmological
information content is boosted, relative to the standard approach of
\citet{Tegmark1997}.
\end{abstract}



\section{Introduction}\label{eq:intro}

The number counts of massive galaxy clusters has long been known to
provide strong constraints on the cosmological model, provided one
understands how to map from observed mass proxies to a theoretical
halo mass \citep[for a recent review see][]{Allenetal2011}. Since the
first measurements of the clustering of Abell clusters was performed
in the 70's and early 80's, it has been understood that the clustering
of clusters contains additional vital information about the
cosmological model \citep{HauserPeebles1973,BahcallSoneira1983}. In
particular, these studies were able to show that the clustering of
clusters was stronger than that of the galaxies. This quickly lead to
the realization that galaxies and clusters could not both be unbiased
tracers of the mass distribution \citep{Kaiser1984}. One of the major
attractions of the cold dark matter (hereafter, CDM) framework, is
that `biased' clustering naturally emerges within it. In Kaiser's
seminal work, he showed that the peaks and troughs of a Gaussian
random field were correlated more strongly than the correlation
function of the unconstrained field. Under the assumption that the
Abell clusters formed out of the high peaks of a Gaussian Random Field
one would then expect the Abell clusters to be more strongly
correlated than galaxies. Further theoretical support comes from the
excursion set formalism, which showed that initially overdense patches
of a CDM universe would collapse to form dark matter haloes, and that
these would, in general, be positively biased with respect to the
underlying matter \citep{ColeKaiser1989,MoWhite1996,ShethTormen1999}.
One important consequence of these developments was that in the mid
90s, it was also realized that, if one combined measurements of the
clustering of clusters with measurements of their abundances, one
could break the degeneracies in cosmological parameters that were
inherent in one single method
\citep{MoJingWhite1996,MajumdarMohr2004,LimaHu2004,LimaHu2005,
  Oguri2009,Cunhaetal2010,SmithMarian2011,OguriTakada2011}.

Some notable measurements of the clustering of clusters are: in the
X-rays, initial measurements of the cluster correlation function for
ROSAT data were performed by \citet{Romeretal1994}, and were later
improved upon by \citet{Collinsetal2000} using the 344 clusters in the
REFLEX survey \citep[see also][for results from the XBACS
  survey]{Abadietal1998}. In the optical, cluster samples tend to be
orders of magnitude larger \citep[for a review of early results
  see][]{Bahcall1988}. The APM galaxy survey was able to identify
several hundred clusters for which clustering was computed
\citep{Daltonetal1992,Milleretal2001a,Milleretal2001b}. In the past
decade, the Sloan Digital Sky Survey (SDSS) has produced, by far, the
largest homogeneous sample of optical clusters: the MaxBCG sample
whose clusters are detected via the `red sequence' cluster detection
method in multi-band imaging data \citep{Koesteretal2007}. This sample
contains 13,823 clusters with velocity dispersions $\gtrsim$400 ${\rm
  km\, s}^{-1}$ and covers an area of $\gtrsim 7000$ square
degrees. The cluster correlation functions were explored by
\citet{Bahcalletal2003} and \citet{Estradaetal2009}, and the power
spectrum analysis was performed by \citet{Huetsi2010}.

In the future, X-ray cluster surveys, such as eROSITA, should produce
homogeneous cluster samples with numbers of clusters on the order of
$\sim$100,000 \citep{Pillepichetal2012}. Deep multi-band optical
surveys such as the Dark Energy Survey\footnote{\tt
  www.darkenergysurvey.org} should also produce tens of thousands of
high signal-to-noise ratio (hereafter, $\SN$) clusters. The question
then arises: how should one perform an optimal measurement of the
clustering of galaxy clusters?  In a series of recent theoretical
studies \citep{Seljaketal2009,Hamausetal2010}, it was claimed that if
galaxy clusters were weighted by a function with a linear dependence
on mass, then the shot-noise on cluster power spectra measurements
would be significantly reduced, hence yielding improved cosmological
information. A complex study by \citet{Caietal2011}, proposed that the
optimal way to reconstruct the mass distribution from a set of
clusters was to weight the galaxy clusters by some combination of
their mass and bias. In the limit of a low mass detection threshold
for the clusters, these works lead to the conclusion that weighting by
mass results in a maximal $\SN$ measurement of the matter density
fluctuations. However, the caveat to the above analysis was that
neither work directly demonstrated that the $\SN$ for cluster power
spectra would be maximized. In this paper we shall directly perform
this task. As we shall show, our analysis generalizes the galaxy power
spectrum methods developed by \citet[][hereafter FKP]{Feldmanetal1994}
and \citet[][hereafter PVP]{Percivaletal2004a}.

This paper is broken down as follows: In \S\ref{sec:survey} we
overview the specifications of a cluster survey, the construction of
the density field of clusters, and its basic statistical properties.
In \S\ref{sec:twopoint} we detail the estimators of the two-point
correlation function and the power spectrum. In \S\ref{sec:fourpoint}
we write down the covariance matrix of the power spectrum estimator in
the most general form and the Gaussian limit.  In \S\ref{sec:optimal}
we provide details of the derivation of the optimal weighting scheme.
In \S\ref{sec:case} we compare various weighting schemes with our
optimal weights for the cases of volume limited and flux-limited
cluster surveys. We also present a new expression for the Fisher
information matrix, which may be used for predicting the cosmological
information content of optimally weighted measurements of the cluster
power spectrum.  Finally, in \S\ref{sec:conclusions} we summarize our
findings and conclude.


\section{Survey specifications and the $\Fc$--field}\label{sec:survey}

\subsection{A generic cluster survey}

Let us begin by defining our fiducial cluster survey: suppose that we
have observed $N$ clusters and to the $i$th cluster we assign a mass
$M_i$, redshift $z_i$ and angular position on the sky
$\bm\Omega_i=\bm\Omega(\theta_i,\phi_i)$. The cluster selection
function depends on both position and cluster mass and in general, is
a complex function of the survey flux limit, and the cluster detection
procedure \citep[for an example of the complexities involved in
  computing this for the eROSITA mission
  see][]{Pillepichetal2012}. However, it may be simplified in the
following ways. Firstly, provided the flux-limit is homogeneous across
the survey area, the angular and radial parts of the selection
function are separable:
\be \Theta(\bx|M) = \Theta(\bm\Omega) \Theta(\chi|M), \ee
where $\chi=\chi(z)$ is the radial comoving geodesic distance to redshift
$z$.  The angular selection function may be written:
\be \Theta(\bm\Omega) = 
\left\{
\begin{array}{cc}
1 \ ; & [{\bm\Omega} \in {\bm\Omega}_{\mu} ] \\
0 \ ; & [\rm otherwise] 
\end{array} \ ,
\right.
\ee
where $\bm\Omega_{\mu}$ defines the survey mask. The radial selection
function may be written:
\be \Theta(\chi|M) = \left\{
\begin{array}{cc}
1 \ ; & [{\chi} \in \chi_{\rm max}(M) ] \\
0 \ ; & [\rm otherwise] 
\end{array} \ ,
\right.
\ee
where $\chi_{\rm max}(M)$ is the maximum comoving geodesic distance
out to which a cluster of mass $M$ could have been detected. This last
relation may be inverted to obtain a very useful relation, which is
the minimum detectable cluster mass at radial position $\chi(z)$ in
the survey.  We shall denote this quantity as $\Mlim(\chi)$. 

Secondly, if the survey is volume limited, the minimum detectable mass
is independent of position and we may write:
\be \Theta^{\rm VL}(\chi|M) = \Theta^{\rm VL}(\chi)\Theta^{\rm VL}(M) \ ,
\ee
where 
\be \Theta^{\rm VL}(\chi|M) = \left\{
\begin{array}{cc}
1 \ ; & [{\chi} \le \chi_{\rm max} ] \\
0 \ ; & [\rm otherwise] 
\end{array} \ 
\right.
\hspace{1cm} \ ; \hspace{1cm}
\Theta^{\rm VL}(M) = \left\{
\begin{array}{cc}
1 \ ; & [M \ge \Mlim ] \\
0 \ ; & [\rm otherwise] 
\end{array} \ .
\right.
\ee

The survey volume may now be defined as the integral of the selection
function over all space:
\be \Vmu(M) = \int^{\chi_{\rm max}(M)}_{0} \Theta(\bx|M)
dV(\chi,\bm\theta) = \Omega_{\mu} \int_{0}^{\chi_{\rm max}(M)} d\chi 
D^2_A(\chi) ,\ee
where $dV(\chi,\bm\theta)$ is the comoving volume element at position
vector $\bx(\chi(z),\bm\theta)$, $D_{A}(\chi)$ is the comoving angular
diameter distance. For a flat space-time geometry the survey volume
simplifies to,
\be \Vmu(M) = \frac{\Omega_{\mu}}{3} \chi_{\rm max}^3(M) \ .\ee
%
\subsection{The cluster delta expansion}

In general the spatial density distribution of clusters, per unit
mass, at position $\bx(\chi,\bm\Omega)$ may be written as a sum over
Dirac delta functions:
\be 
n_{\c}(M,\bx)=\sum_{i=1}^{N}\delta^{\rm D}(M-M_i)\delta^{\rm D}(\bx-\bx_i) \ .
\label{eq:clusexp} 
\ee 
If the selection function is inhomogeneous, then the mean density of
clusters varies spatially over the survey. Next, in analogy with PVP,
we define a field $\Fc$, which is related to the over-density of
clusters. This can be written:
\be 
\Fc(\bx) = \int dM \frac{w(\bx,M)}{\sqrt{A}}\Theta(\bx | M)
\left[n_{\c}(\bx,M)-\alpha n_s(\bx,M)\right] 
\label{eq:clusterden} \ ,
\ee
where $n_s(\bx,M)$ represents the number density of clusters in a mock
sample that has no intrinsic spatial correlations, and whose density
is $1/\alpha$ times that of the true cluster field at that mass. Note
that whilst the field $n_s(\bx,M)$ has no intrinsic spatial
correlations it does possess a spectrum of masses, which is closely
related to the mass spectrum of the field
$\left<n_{\c}(\bx,M)\right>$.  The choice for the normalization
parameter $A$ will be given later.  The quantity $w(\bx,M)$ denotes a
weight function that, in general, may depend on both the spatial
position and mass of the cluster. It is this quantity that we shall
aim to determine in an optimal way.

\subsection{Statistical properties of the cluster density field}
Determination of the optimal weight function will require statistical
analysis on the field $\Fc$, therefore we will now introduce the
necessary tools.  As a simple example let us compute the ensemble
average value of the field $\Fc$, which can be written,
\be 
\left< \Fc(\bx)\right>  = \int dM \frac{w(\bx,M)}{\sqrt{A}} \Theta(\bx | M)
\left[\left<n_{\c}(\bx,M)\right>-\alpha \left<n_s(\bx,M)\right>\right]\ , 
\label{eq:Fensemb}
\ee
where the angled brackets denote an ensemble average in the following
sense:
\be \left< B(\left\{\bx_i,M_i\right\}) \right>  \equiv 
\int \dx_1 \dots \dx_N \dM_1 \dots \dM_N 
p(\bx_1,\dots,\bx_N,M_1,\dots,M_N)
B(\left\{\bx_i,M_i\right\}) \ .
\ee
in the above $p(\bx_1,\dots,\bx_N,M_1,\dots,M_N)$ is the $N$-point
joint probability distribution for the $N$ clusters being located at
the set of spatial positions $\{\bx_i\}$ and having the set of masses
$\{M_i\}$. Thus, the first expectation on the right-hand side of
\Eqn{eq:Fensemb} can be written as:
\be  
\left< n_{\c}(\bx,M) \right>  =  \int \prod_{l=1}^{N} \left\{\dx_l \dM_l\right\}
p(\bx_1,\dots,\bx_N,M_1,\dots,M_N) \sum_{i=1}^{N}\delta^{\rm D}(\bx-\bx_i)\delta^{\rm D}(M-M_i) 
=  \sum_{i=1}^{N} p(\bx,M) = N p(\bx,M) \ .
\ee
On the first line we inserted the expansion of the cluster density
field from \Eqn{eq:clusexp} and to obtain the second we integrated
over the sum of Dirac delta functions. The quantity $p(\bx,M)$ can be
written in a more transparent way:
\be 
p(\bx,M) \equiv p(\bx) p(M) 
= \frac{1}{V}\frac{\nbar(M)}{N/V} = \frac{\nbar(M)}{N} \ ,
\ee
where $\bar{n}(M)$ is the intrinsic number density of clusters per
unit mass -- i.e. the dark matter halo mass function.

Turning to the second expectation value, we note that the only
difference between $\left<n_{\c}(\bx,M) \right>$ and
$\left<n_s(\bx,M)\right>$ is the artificially increased space density
of clusters and the absence of any intrinsic clustering.  Hence, we
also have,
\be \alpha\left< n_{s}(\bx,M) \right> = \bar{n}(M)\ .\ee
Putting this all together, we arrive at the result:
\be \left<\Fc(\bx)\right>=0 \ .\ee
Hence the $\Fc$--field, like the over-density field of matter, is a
mean-zero field.

Note that we have neglected to take into account the statistical
properties of obtaining the $N$ clusters in the survey volume. In what
follows we shall assume that the survey volumes are sufficiently large
that this may be essentially treated as a deterministic
quantity. However, it can be taken into account
\citep[e.g. see][]{ShethLemson1999,SmithWatts2005,Smith2009}.

\section{Clustering Estimators}\label{sec:twopoint}
\subsection{An estimator for the two-point correlation function}

The two-point correlation function of the field $\Fc$ can be computed 
directly as:
\ba
\left<\Fc(\bx_1)\Fc(\bx_2)\right> &\hspace{-0.2cm} =\hspace{-0.2cm} & \frac{1}{A}
\int dM_1 dM_2 w(\bx_1, M_1)w(\bx_2, M_2)\Theta(\bx_1|M_1)\Theta(\bx_2| M_2)
\left[\frac{}{}\hspace{-0.1cm}\left<n_\c(\bx_1, M_1) n_\c(\bx_2, M_2)\right>\right. \nn \\
& & \hspace{-0.69cm}\left.\frac{}{}-\alpha\left<n_\c(\bx_1, M_1) n_s(\bx_2, M_2)\right> 
-\alpha\left<n_\c(\bx_2, M_2)n_s(\bx_1, M_1)\right> 
+\alpha^2\left<n_s(\bx_1, M_1) n_s(\bx_2, M_2)\right>
\right] \label{eq:FF} \ .
\ea
The calculation of the above terms is detailed in the
appendix~\ref{app:AI}. On substituting the results of
\Eqns{eq:ncnc}{eq:ncns}, we write
\ba
\xi_{\Fc}(\bx_1,\bx_2)  & = & \frac{1}{A}\int dM_1 dM_2 w(\bx_1, M_1)
w(\bx_2,M_2)\nbar(M_1)\nbar(M_2)\Theta(\bx_1|M_1) \Theta(\bx_2|M_2)
\xi^\c(\bx_1, \bx_2, M_1, M_2) \nn \\
& & \hspace{-0.6cm} + \;\;\frac{(1+\alpha)}{A}\int dM w^2(\bx_1, M)\nbar(M)\Theta(\bx_1|M)
\delta^{\rm D}(\bx_1-\bx_2)
\label{eq:FcorrClust} \ ,
\ea
where we defined
$\xi_{\Fc}(\bx_1,\bx_2)\equiv\left<\Fc(\br_1)\Fc(\br_2)\right>$. Note
that in the second line of the above equation we have benefited from
the identity: $\Theta^2(\bx|M)=\Theta(\bx|M)$. If we assume that the
cluster density field is some local function of the underlying dark
matter density
\citep{FryGaztanaga1993,MoWhite1996,Moetal1997,Smithetal2007}, the
cross-correlation function of clusters of masses $M_1$ and $M_2$, at
leading order, can be written:
\be 
\xi^\c(|\bx_1-\bx_2|,M_1, M_2) = b(M_1) b(M_2)\xi(|\bx_1-\bx_2|) \ ,
\label{eq:linbias}
\ee
where $\xi(r)$ is the correlation of the underlying matter
fluctuations. On inserting this relation into \Eqn{eq:FcorrClust} we
find:
\be
\xi_{\Fc}(\bx_1,\bx_2)   =  \G_{(1,1)}(\bx_1)\G_{(1,1)}(\bx_2)
\xi(|\bx_1-\bx_2|) +(1+\alpha)\G_{(2,0)}(\bx_1) \delta^{D}(\bx_1-\bx_2)
\label{eq:FcorrClustLinb} \ ,
\ee
where we have defined the weighted selection function 
\be
\G_{(l,m)}(\bx) \equiv \frac{1}{A^{l/2}}\int dM \nbar(M) b^m(M) 
w^l(\bx, M) \Theta(\bx|M) \ .
\label{eq:winReal} 
\ee
\noindent One possible estimator for the matter correlation function from the
$\Fc$ field is therefore:
\be \hat{\xi}_{\Fc}(\bx) \equiv \int \dx' \Fc(\bx)\Fc(\bx+\bx') \ \ \ ; \ \ \ (\bx\ne{\bf 0}) \ .\ee 
If we now compute the expectation of this estimator we find:
\be
\left<\hat{\xi}_{\Fc}(\bx)\right> = \xi(\bx) \int \dx'\, \G_{(1,1)}(\bx)
\G_{(1,1)}(\bx+\bx') \ \ \ ; \ \ \ (\bx\ne{\bf 0})
\ .
\ee
Although $\hat{\xi}_{\Fc}(\bx)$ is not an unbiased estimator of the
matter correlation function, we may construct one that is:
\be
\hat{\xi}_0(\bx) \equiv \hat{\xi}_{\Fc}(\bx) /\Xi_0(\bx) \hspace{0.3cm} ; 
\hspace{0.3cm}
\Xi_0(\bx)\equiv \int \dx'\, \G_{(1,1)}(\bx)\G_{(1,1)}(\bx+\bx') \ .
\label{eq:xiest}
\ee
%
\subsection{An estimator for the cluster power spectrum}
We may now compute the Fourier space equivalent of the two-point
correlation function, the power spectrum. In what follows we shall
adopt the following Fourier transform conventions: 
\be
\tilde{B}(\bk) =  \int \dx B(\bx)\ex^{i\bk\cdot\bx}\hspace{0.5cm} 
\Leftrightarrow\hspace{0.5cm} 
B(\bx) =  \int \frac{\dk}{(2\pi)^3} {\tilde B}(\bk) \ex^{-i\bk\cdot\bx} \ .
\nn \ee
We also define the power spectrum, $P_B(k)$, of any infinite
statistically homogeneous random field $\tilde B(\bk)$ to be:
\be 
\left<{\tilde B}(\bk){\tilde B}(\bk')\right> \equiv
(2\pi)^3\delta^{\rm D}(\bk+\bk')P_B(\bk) \ . 
\nn \ee
Note that if the field $B$ were statistically isotropic, the power
spectrum would simply be a function of the scalar $k$.  With the above
definitions in hand, the covariance of the Fourier modes of the
cluster field $\Fc$ can be written:
\ba 
\left<\Fcf(\bk_1)\Fcf(\bk_2)\right> & = & \frac{1}{A}
\int \prod_{i=1}^2 \left\{\dx_i dM_i w(\bx_i,M_i)\nbar(M_i)\Theta(\bx_i|M_i)
\ex^{\bk_i\cdot\bx_i}\right\}\xi^\c(\bx_1,\bx_2,M_1,M_2)\nn \\
& & 
+(1+\alpha)\int \dx \,dM \nbar(M)\Theta(\bx|M)w^2(\bx, M)\ex^{i(\bk_1+\bk_2)\cdot\bx}  \ . 
\label{eq:PkFc}
\ea
For any infinite homogeneous random field, the two-point correlation
function and the power spectrum form a Fourier pair, hence we may
write:
\be 
\xi^\c(\bx_1,\bx_2,M_1,M_2)=
\int \frac{\dk }{(2\pi)^3}P_c(\bk, M_1, M_2)\ex^{-i\bk\cdot(\bx_1-\bx_2)}\ .
\nn \ee
If we assume the linear biasing relation of \Eqn{eq:linbias}, then the
cross-power spectrum of clusters of different masses $M_1$ and $M_2$,
at leading order, can be written
\be P_c(\bk, M_1, M_2) = b(M_1) b(M_2) P(\bk) \ ,
\ee
where $P(\bk)$ is the matter power spectrum. On using these
expressions in \Eqn{eq:PkFc}, and considering the case $\bk_1=-\bk_2$ we
find:
\be \left<|\Fcf(\bk)|^2\right> =  
\int \frac{\dq }{(2\pi)^3}P(\bq) \left|\Gf_{(1,1)}(\bk-\bq)\right|^2 + P_{\rm shot} \ ,
\label{eq:Pf} \ee
where in the above expression $\Gf_{(l,m)}(\bk)$ is the
Fourier transform of the weighted survey selection function from
\Eqn{eq:winReal} and we have defined the shot-noise term as:
\be
P_{\rm shot} \equiv (1+\alpha) \Gf_{(2,0)}({\bf 0}) \ .
\label{eq:Pshot}
\ee
Just as in the case of galaxies \citep{Feldmanetal1994}, the
expectation of the square amplitude of the Fourier modes of $\Fc$ is
given by the convolution of the matter power spectrum with the modulus
square of the Fourier modes of the survey window function, plus a
constant shot noise.

In the limit that the survey volume is large, the functions
$\Gf_{(l,m)}(\bk)$ will be very narrowly peaked around $\bk=0$.
Provided the matter power spectrum is a smoothly varying function of
scale, the window functions $\Gf_{(l,m)}(\bk)$ take on Dirac
delta-function-like behaviour. Thus, in the large-survey volume limit,
the first term on the right-hand side of \Eqn{eq:Pf} becomes:
\be
\int \frac{\dq}{(2\pi)^3}P(\bq) \left|\Gf_{(1,1)}(\bk-\bq)\right|^2 
\approx  P(\bk) \int \frac{\dq}{(2\pi)^3} \left|\Gf_{(1,1)}(\bk-\bq)\right|^2 \ .
\ee
Let us focus on the integral factor on the right-hand-side of the
above expression. Transforming the variable $\bk-\bq \rightarrow \bq'$
and using Parseval's theorem, as well as \Eqn{eq:winReal}, we find:
\be
\int \frac{\dq}{(2\pi)^3} \left|\Gf_{(1,1)}(\bk-\bq)\right|^2 
= \int \frac{\dq'}{(2\pi)^3} \left|\Gf_{(1,1)}(\bq')\right|^2 
= \int \dx \left[\G_{(1,1)}(\bx)\right]^2 =  \frac{1}{A}\int \dx \left[\int dM  
\nbar(M) b(M) w(\bx, M) \Theta(\bx | M)\right]^2\ . 
\label{eq:Gconv}\ee
Recall that we have not yet specified the parameter $A$, let us now
define it to be:
\be 
A \equiv\int\dx \left[\G_{(1,1)}(\bx)\right]^2 = \int \dx
\left[\int dM \nbar(M) b(M) w(\bx, M) \Theta(\bx|M)\right]^2 \ .
\label{eq:Norm}
\ee
With this choice of normalization, \Eqn{eq:Gconv} is simply
unity. Hence, for the case of large homogeneous survey volumes, an
unbiased estimator for the dark matter power spectrum is:
\be \hat{P}(\bk) \approx |\Fcf(\bk)|^2 - P_{\rm shot} \ .
\label{eq:Pest}
\ee
The above estimator is for the power spectrum at a particular mode,
whereas we are more interested in a band-power estimate of the power
spectrum. Thus, our final estimator is:
\be
\hat{\overline{P}}(k_i) = \frac{1}{V_i}\int_{V_i} \dk \hat{P}(\bk) = \frac{1}{V_i}\int_{V_i} 
\dk |\Fcf(\bk)|^2 -P_{\rm shot}\ ,
\label{eq:Pest2}
\ee
where in the above we have summed all modes over a shell in $k$-space
of thickness $\Delta k$, i.e.
\be 
V_{i} \equiv \int_{V_i} \dk 
= 4\pi \int^{k_i+\Delta k/2}_{k_i-\Delta k/2} k^2 dk 
= 4\pi k_i^2 \Delta k
\left[1+\frac{1}{12}\left(\frac{\Delta k}{k_i}\right)^2\right] \ .
\nn \ee
If the survey window function possesses small-scale structure, then
the matter power spectrum can only safely be recovered by
deconvolution of the window function
$\left|\Gf_{(1,1)}(\bk)\right|^2$, or alternatively one must convolve
theory predictions with the window function. Otherwise, \Eqn{eq:Pest}
is a biased estimator.
\section{Statistical fluctuations in the cluster power spectrum}
\label{sec:fourpoint}
In order to obtain the optimal estimator we need to know how the $\SN$
varies with the shape of the weight function $w(\bx,M)$. Hence, we
need to understand the noise properties of our power spectrum
estimator.

\subsection{The covariance of the cluster power spectrum estimator}
In general, the covariance matrix of the band-power spectrum estimator
can be written as:
\be 
{\rm Cov}\!\left[\hat{\overline{P}}(k_i),\hat{\overline{P}}(k_j)\right] 
\equiv \left< \hat{\overline{P}}(k_i)\hat{\overline{P}}(k_j)\right> -
\langle\hat{\overline{P}}(k_i)\rangle\langle\hat{\overline{P}}(k_j)\rangle
= \frac{1}{V_{i}}\int_{V_j} \dk_1 \frac{1}{V_j}\int_{V_j} \dk_2 
{\rm Cov}\!\left[\hat{P}(\bk_1),\hat{P}(\bk_2)\right] \ .
\nn \ee
In Appendix~\ref{app:AII.1} we show that the latter term can be
written as:
\be
{\rm Cov}\!\left[\hat{P}(\bk_1),\hat{P}(\bk_2)\right] \approx
{\rm Cov}\!\left[|\Fcf(\bk_1)|^2,|\Fcf(\bk_2)|^2\right] = 
\left<|\Fcf(\bk_1)|^2|\Fcf(\bk_2)|^2\right>-
\left<|\Fcf(\bk_1)|^2\right> \left<|\Fcf(\bk_2)|^2 \right> 
\label{eq:CovPest1}
\ee
As detailed in Appendix~\ref{app:AII.2}, the covariance can be written
as:
\ba
{\rm Cov}\!\left[|\Fcf(\bk_1)|^2,|\Fcf(\bk_2)|^2\right] 
\hspace{-0.2cm} & = &\hspace{-0.2cm} \frac{1}{A^2}
\int \dk_3 \,\dk_4 \delta^{\rm D}(\bk_1+\bk_3) \delta^{\rm D}(\bk_2+\bk_4)
\int \dx_1 \dots \dx_4 \,\ex^{i\bk_1\cdot\bx_1+\dots+\bk_4\cdot\bx_4}  \nn \\
& \times & 
\left[\left<\Fc(\bx_1)\dots \Fc(\bx_4)\right> - \left<\Fc(\bx_1) \Fc(\bx_3)\right>
\left<\Fc(\bx_2) \Fc(\bx_4)\right> \right] \ ,
\label{eq:CovF1}
\ea 
where the last term in the equation above is given by \Eqn{eq:CovF0}.
In Appendix~\ref{app:AII.3} we provide a general relation for the
covariance of the $\Fc$ power spectrum in terms of Fourier-space
quantities. These results generalize the expressions for the
covariance matrix of the cluster power spectrum presented in
\citep{Smith2009}, extending that work to the case of finite survey
geometry and an arbitrary weighting scheme that depends on mass and
position.

Under the assumption that the matter density field is Gaussianly
distributed, all connected correlation functions beyond two-point ones
vanish (e.g. $\zeta$ and $\eta$ from \Eqn{eq:def_high}). The general
expression given by \Eqn{eq:CovPowerGen} simplifies to:
\ba
{\rm Cov}\!\left[|\Fcf(\bk_1)|^2,|\Fcf(\bk_2)|^2\right] 
& = &
\left|\int \frac{\dq_1}{(2\pi)^3}
P(\bq_1)\Gf_{(1,1)}(\bk_1-\bq_1)\Gf_{(1,1)}(\bk_2+\bq_1)
+(1+\alpha)\Gf_{(2,0)}(\bk_1+\bk_2)\right|^2
\nn \\ &  + & 
\left|\int \frac{\dq_1}{(2\pi)^3} P(\bq_1)\Gf_{(1,1)}(\bk_1-\bq_1)\Gf_{(1,1)}(-\bk_2+\bq_1) 
+(1+\alpha)\Gf_{(2,0)}(\bk_1-\bk_2)\right|^2
\nn \\ & + & 
\int \frac{\dq_1}{(2\pi)^3} P(\bq_1) \left\{
\left|\Gf_{(2,1)}(\bk_1+\bk_2-\bq_1)\right|^2+\left|\Gf_{(2,1)}(\bq_1)\right|^2
+\left|\Gf_{(2,1)}(\bk_1-\bk_2-\bq_1)\right|^2\right.
\nn \\ & + & 
\Gf_{(3,1)}(\bk_1-\bq_1)\Gf_{(1,1)}(-\bk_1+\bq_1)
+\Gf_{(3,1)}(\bk_2-\bq_1)\Gf_{(1,1)}(-\bk_2+\bq_1)
\nn \\
& + & \left.\Gf_{(3,1)}(-\bk_1-\bq_1)\Gf_{(1,1)}(\bk_1+\bq_1)
+\Gf_{(3,1)}(-\bk_2-\bq_1)\Gf_{(1,1)}(\bk_2+\bq_1)
\right\}+ (1+\alpha^3)\Gf_{(4,0)}(0) \ . 
\label{eq:CovGauss1}
\ea
To further proceed with the power spectrum covariance, let us
introduce the functions:
\be
\Q^{(i_1, i_2)}_{(j_1, j_2)}(\bx) \equiv \G_{(i_1, j_1)}(\bx) \G_{(i_2, j_2)}(\bx) \ ;
\hspace{0.5cm} \mbox{and} \hspace{0.2cm} \Q^{(i)}_{(j)}(\bx) \equiv \G_{(i, j)}(\bx) \ .
\label{eq:Qdef}
\ee
The Fourier transform of $\Q^{(i_1, i_2)}_{(j_1, j_2)}(\bx)$ is
obtained through the convolution theorem:
\be 
\Qf^{(i_1, i_2)}_{(j_1, j_2)}(\bk) \equiv \int \frac{\dq}{(2\pi)^3}
\Gf_{(i_1,j_1)}(\bq) \Gf_{(i_2, j_2)}(\bk-\bq) \ .
\label{eq:QFourier}
\ee
Consider the case where the survey volume is large and hence the
$\Gf_{j,k}(\bk)$ functions are all very narrowly peaked around
$\bk=0$. With this assumption and using \Eqns{eq:Qdef}{eq:QFourier},
we express \Eqn{eq:CovGauss1} more compactly as:
\ba
{\rm Cov}\!\left[|\Fcf(\bk_1)|^2,|\Fcf(\bk_2)|^2\right] 
\hspace{-0.2cm} & \approx & \hspace{-0.2cm}
\left| P(\bk_1) \Qf^{(1,1)}_{(1,1)}(\bk_1\hspace{-0.05cm}+\hspace{-0.05cm}\bk_2)
+(1+\alpha)\Qf^{(2)}_{(0)}(\bk_1\hspace{-0.05cm}+\hspace{-0.05cm}\bk_2)\right|^2 
\hspace{-0.1cm}+\hspace{-0.1cm}\left|P(\bk_1)\Qf^{(1,1)}_{(1,1)}
(\bk_1\hspace{-0.05cm}-\hspace{-0.05cm}\bk_2)+(1+\alpha)
\Qf^{(2)}_{(0)}(\bk_1\hspace{-0.05cm}-\hspace{-0.05cm}\bk_2)\right|^2 \nn \\
& & \hspace{-3cm}
+ \Qf^{(2,2)}_{(1,1)}({\bf 0})\left[\frac{}{}\hspace{-0.1cm}
P(\bk_1\hspace{-0.05cm}+\hspace{-0.05cm}\bk_2)+
P(\bk_1\hspace{-0.05cm}-\hspace{-0.05cm}\bk_2)\right] 
+ \Qf^{(3,1)}_{(1,1)}({\bf 0})\left[\frac{}{}\hspace{-0.1cm} P(\bk_1)+P(\bk_2)+P(-\bk_1)
+P(-\bk_2)\right]+ (1+\alpha^3)\Qf^{(4)}_{(0)}({\bf 0}) \ . 
\label{eq:CovGauss2}
\ea
Notice that the equation above is invariant under $\bk_i\rightarrow
-\bk_i,\, i\in(1,2)$. We shall use this property next, when evaluating
the bin-averaged covariance of the $\Fc$ power spectra: 
\ba
{\rm Cov}\!\left[|\Fcf(k_i)|^2,|\Fcf(k_j)|^2\right] \hspace{-0.2cm} & \approx & \hspace{-0.2cm}
2 \overline{P}^2(k_i) \hspace{-0.1cm}\int_{V_i} \hspace{-0.2cm}\frac{\dk_1}{V_i} 
\hspace{-0.1cm} \int_{V_j} \hspace{-0.2cm}\frac{\dk_2}{V_j} 
\Qf^{(1,1)}_{(1,1)}(\bk_1+\bk_2)\Qf^{(1,1)}_{(1,1)}(-\bk_1-\bk_2)
+ \,4 (1+\alpha)\overline{P}(k_i) \nn \\  &  &  \hspace{-2cm}
\times  \int_{V_i} \hspace{-0.1cm}
\frac{\dk_1}{V_i} \hspace{-0.1cm}\int_{V_j}\hspace{-0.1cm} \frac{\dk_2}{V_j} 
 \Qf^{(1,1)}_{(1,1)}(\bk_1+\bk_2)\Qf^{(2)}_{(0)}(-\bk_1-\bk_2)
+ 2(1+\alpha)^2\int_{V_i} \hspace{-0.1cm}\frac{\dk_1}{V_i}
\int_{V_j}\hspace{-0.1cm} \frac{\dk_2}{V_j} 
\Qf^{(2)}_{(0)}(\bk_1+\bk_2)\Qf^{(2)}_{(0)}(-\bk_1-\bk_2)
\nn \\ & & \hspace{-2cm}
+\: 2 \Qf^{(2,2)}_{(1,1)}({\bf 0}) \overline{P}[k_i,k_j]  +
2\Qf^{(3,1)}_{(1,1)}({\bf 0}) \left[{\overline P}(k_i) +\overline{P}(k_j) \right]
+ (1+\alpha^3)\Qf^{(4)}_{(0)}({\bf 0}) \ ,
\label{eq:CovGauss3}
\ea
where we defined:
\be \overline{P}[k_i,k_j] \equiv \int_{V_i} \frac{\dk_1}{V_i}
\int_{V_j} \frac{\dk_2}{V_j} P(\bk_1+\bk_2) \ .
\nn 
\ee
To get \Eqn{eq:CovGauss3}, we expanded the modulus-squared terms in
\Eqn{eq:CovGauss2} and we assumed that the $k$-space shells are
sufficiently narrow that the power spectrum can be considered constant
over the shell. In Appendix~\ref{app:AII.4} we show how to compute the
$k$-space-shell averages of products of the $\Qf$ functions. Using the
result in \Eqn{eq:result}, we write the covariance matrix of the power
spectra of the field $\Fc$ from \Eqn{eq:CovGauss3} as:
\ba
{\rm Cov}\!\left[|\Fc(k_i)|^2,|\Fc(k_j)|^2\right] & \approx & 
\frac{2(2\pi)^3}{V_i} \left[\overline{P}^2(k_i)\,\Xi_{(1,1|1,1)}^{(1,1|1,1)}(0)
+ 2(1+\alpha)\overline{P}(k_i) \,\Xi_{(1,1|0)}^{(1,1|2)}(0)
+ (1+\alpha)^2 \,\Xi_{(0|0)}^{(2|2)}(0) \right]\delta^{K}_{i,j}\nn \\
& + & 2\Qf_{(1,1)}^{(2,2)}({0}) \overline{P}[k_i,k_j]  + 2\Qf^{(3,1)}_{(1,1)}({0})
\left[{\overline P}(k_i) +\overline{P}(k_j) \right] + (1+\alpha^3)\Qf^{(4)}_{(0)}({0}) \ ,
\label{eq:CovGauss4}
\ea
where the correlation function $\Xi^{(i_1, i_2|l_1, l_2)}_{(j_1, j_2|
  m_1, m_2)}$ of the weighted survey windows is defined by
\Eqn{eq:Xidef}. Finally, on taking the limit that $\nbarh \Vu\gg 1$,
the last three terms on the right-hand side of the above equation will
be sub-dominant \citep{Smith2009}. Using the expressions in
\Eqn{eq:Xiused}, we arrive at the simplified form:
\be
{\rm Cov}\!\left[|\Fc(k_i)|^2,|\Fc(k_j)|^2\right] \approx
\frac{2(2\pi)^3}{V_{i}} \overline{P}^2(k_i) \int \dx   \left\{ 
\left[\G_{(1,1)}(\bx)\right]^2 + \frac{(1+\alpha)}{\overline{P}(k_i)}\G_{(2,0)}(\bx)\right\}^2
\delta^{K}_{i,j} \ .
\label{eq:CovGauss5}
\ee
Note that the weights derived in the next section are only optimal
under some important approximations: the underlying matter density is
Gaussian; the sampling fluctuations for a given realization of the
cluster field are small; the survey volume is large and
homogeneous. In future work it will be interesting to explore how the
optimal weights change as each of these assumptions is gradually
relaxed.

At this point it is also worth comparing our analysis with that of FKP
and PVP.  Our expressions for the correlation function \Eqn{eq:xiest}
and the power spectrum \Eqn{eq:Pest} are analogous to those found in
PVP for the luminosity dependence of the galaxy bias. However, our
derivation is different, particularly in the way in which we treat the
statistical properties of the cluster field, which we believe to be
more general and transparent. Thus, our method establishes a clear
formalism for evaluating how the optimal weights change when the
above-mentioned assumptions are relaxed, which will be very useful to
future studies of optimal power spectra estimators. Although our choice
of the normalization parameter $A$ is different from PVP, it is of no
relevance until we derive $w(\br,M)$.  Indeed, in the next section we
will show that, once the optimal weights are selected, their and our
choice for the field $\Fc$ turn out to be equivalent.

Comparing to FKP, we also note that our derivation of the statistical
fluctuations in the power and the approximations that we make in order
to obtain a diagonal covariance matrix, are more rigorous and
transparent than in the study of FKP. For instance, the original FKP
derivation makes use of Parseval's theorem to transform from a
band-power $k$-space integral to an integral over the entirety of real
space -- that is they effectively go from \Eqn{eq:CovGauss3} to
\Eqn{eq:CovGauss5} using Parseval's theorem. Owing to the fact that
the band-power averages do not extend over the whole of k-space,
indeed they may be limited to lie only in a very narrow shell, the
application of Parseval's theorem appears incorrect. As a minor point,
their derivation also misses a factor of 2, which arises due to the
Hermitian nature of the Fourier modes. This however plays no role in
the derivation of the optimal weights.
\section{An optimal weighting scheme}
\label{sec:optimal}
\subsection{Optimal weights as a functional problem}
Our aim is to find the optimal weighting scheme that will maximize the
$\SN$ ratio on a given band power estimate of the cluster power
spectrum. To begin, note that maximizing the $\SN$ ratio is equivalent
to minimizing its inverse, the noise-over-signal ratio $\NS$. This can
be expressed as:
\be 
F[w(\bx, M)]\equiv\frac{\sigma^2_{P}(k_i)}{ \overline{P}^2(k_i)} 
=\frac{2(2\pi)^3}{V_{i}} \int \dx   
\left\{ \left[\G_{(1,1)}(\bx)\right]^2 + c \, \G_{(2,0)}(\bx)\right\}^2 \ ,
\label{eq:Func1}
\ee
where we have defined the constant
\be
c \equiv (1+\alpha)\big/\overline{P}(k_i) \ .
\label{eq:def_c}
\ee
In the above expression we have written the quantity $F[w]$ as a
functional of the weights $w(\bx,M)$. The standard way for finding the
optimal weights, is to perform the functional variation of $F$ with
respect to the weights $w$. Operationally, the functional variation of
$F[w]$ may be defined:
\be
\delta F[w(\bx,M)] = F[w(\bx,M)+\delta w(\bx,M)] - F[w(\bx,M)] =
\int \dx dM
\left\{\frac{\delta F}{\delta w(\bx,M)} \right\}\delta w(\bx,M)  \ .
\nn \ee
The extremisation condition is that the functional derivative is
stationary for the optimal weights:
\be \frac{\delta F}{\delta w(\bx, M)}=0\ .
\nn \ee
From \Eqns{eq:winReal}{eq:Norm}, we now note that $F[w]$ is the ratio
of two weight-dependent functionals, since the $\G$ functions contain
weights not only by definition, but also through the normalization
constant $A$:
\be
F[w] = \mathcal{N}[w]\big/\mathcal{D}[w] \ ,
\label{def_Func}
\ee
where we have defined the numerator and denominator functionals:
\begin{align}
\mathcal{N}[w] &\hspace{0.3cm} \equiv \hspace{0.3cm} 
\int \dx \left\{ \left[\,\Gb_{(1,1)}(\bx)\right]^2
+c\, \Gb_{(2, 0)}(\bx)\right\}^2 \label{eq:def_N} \ ; \\
\mathcal{D}[w] & \hspace{0.3cm} \equiv \hspace{0.3cm}  
A^2[w] = \left\{ \int\dx \left[\Gb_{(1, 1)}(\bx)\right]^2\right\}^2 \ .
\label{eq:def_D}
\end{align}
In the above we have ignored the constant $2(2\pi)^3/V_i$ from
\Eqn{eq:Func1}, since it does not play any part in the minimization
process, and we have introduced a scaled version of the survey window
function which is independent of the normalization constant:
\be 
\Gb_{(l,m)} \equiv A^{l/2} \G_{(l,m)}\ .
\label{eq:scaledwin}
\ee
To minimize $F[w]$, we must solve the following functional problem:
\be
\frac{1}{\mathcal{D}[w]}\left[\delta\mathcal{N}[w]-\frac{\mathcal{N}[w]}
{\mathcal{D}[w]} \delta\mathcal{D}[w] \right] = 0 \hspace{0.5cm} \Longleftrightarrow
\hspace{0.5cm} \delta\mathcal{N}[w]-F[w] \delta\mathcal{D}[w] = 0 \ .
\label{eq:minimization}
\ee
In Appendix~\ref{app:A3} we compute the variations of $\mathcal{N}$
and $\mathcal{D}$ with a perturbation $\delta w$. On replacing the
results of \Eqns{eq:fd_N}{eq:fd_D} in \Eqn{eq:minimization} and
dropping all constant terms, we arrive at the general optimal weight
equation:
\be
\left\{\left[\,\Gb_{(1,1)}(\bx)\right]^2 + c\,\Gb_{(2,0)}(\bx)\right\}
\left\{\Gb_{(1,1)}(\bx) b(M) + c\,w(\bx, M)\right\}=\Gb_{(1,1)}(\bx) b(M) \ .
\label{eq:optw1}
\ee
%
\subsection{The optimal weights}
We are now in a position to derive the optimal weights. Consider
\Eqn{eq:optw1}, we examine the scaling with mass and position of the
functions on both sides of the equation. Since the right-hand side is
proportional to a bias term, it follows that the left-hand side must
also obey this proportionality. The first bracket on the left side
does not have any mass dependence. We therefore infer that the
weights' dependence on position and mass must be separable:
\be
w(\bx,M) = b(M) \tilde{w}(\bx) \ .
\label{eq:optw2}
\ee
As an immediate consequence of this separability, the functions $\Gb$
can be written as:
\ba
\Gb_{(1,1)}(\bx)  & = & \tilde{w}(\bx) \int dM \nbar(M) b^2(M) \Theta(\bx | M) \ ; \nn \\
\Gb_{(2,0)}(\bx) & = & \left[\tilde{w}(\bx)\right]^2 \int dM \nbar(M) b^2(M) \Theta(\bx | M) 
 = \tilde{w}(\bx) \Gb_{(1,1)}(\bx) \ .
\nn \ea
Replacing these relations and \Eqn{eq:optw2} in \Eqn{eq:optw1}, we
arrive after a little work at the following solution for the
space-dependent part of the optimal weights:
\be
\tilde{w}(\bx) = 1/\left[c+\int dM \nbar(M) b^2(M) \Theta(\bx | M) \right]\ .
\label{eq:optw3}
\ee
On putting together \Eqns{eq:optw2}{eq:optw3} and substituting back
the constant from \Eqn{eq:def_c}, we write the final expression for
the optimal weights for achieving maximal $\SN$ on a given band-power
estimate of the cluster power spectrum:
\be 
w(\bx,M) = \frac{b(M) }
{(1+\alpha)+\int dM \nbar(M) b^2(M) \Theta(\bx|M) \overline{P}_i} \ .
\label{eq:optw4}
\ee
As in the case of the original FKP weights, we see that the answer is
somewhat circular, in that in order to estimate the cluster power
spectrum optimally, we already need to have some reasonably good
estimate of the underlying matter power spectrum.  Indeed, in order to
fully implement our scheme we are also required to have knowledge of
the functions $b(M)$ and $\nbar(M)$. These two functions are
theoretically very well known for dark matter haloes. They may also be
measured directly from the data -- albeit with noise. One also should
have very good understanding of the selection function
$\Theta(\bx|M)$. The parameter $\alpha$ is determined directly from
the density of the random cluster sample.

We note that \Eqn{eq:optw4} is virtually identical to the result found
by PVP for the case of luminosity-dependent galaxy bias. However, they
provided no analytic proof for their result, but simply proposed a
conjecture for the general weight solution and showed that it
satisfied their weight equation. Our derivation, on the other hand, is
more elegant and easily verified.  Finally, as was pointed out
earlier, once the optimal weights are determined, our choice of field
$\Fc$ and theirs, are virtually equivalent. Their inverse weighting of
the cluster field by the bias, therefore appears to be an unnecessary
step.

Lastly, we generalize \Eqn{eq:optw4} to the case where both $b(M)$ and
$\nbar(M)$ are time dependent quantities. This may be achieved by
simply making them a function of conformal time $\eta$ or equivalently
$\chi$, and also accounting for the presence of the growth factor
$D(\chi)$ in \Eqn{eq:linbias}. If one propagates these transformations
through the entire derivation, then one finds that the weights may be
written:
\be 
w_{\rm OPT}(\bx(\chi,{\bm\Omega}),M) = \frac{D(\chi)b(M,\chi) }
{(1+\alpha)+\int dM \nbar(M,\chi)D^2(\chi)b^2(M,\chi)\Theta(\bx|M)\overline{P}_i} \ .
\label{eq:optimalweight}
\ee
%

\section{Case studies}
\label{sec:case}
We shall now examine how the $\SN$ ratios vary as a function of
minimum cluster mass, for both volume- and flux-limited samples of
clusters. We shall compare the optimal weighting scheme derived in the
previous section with the standard FKP weighting and also the mass
weighting advocated by \citet{Seljaketal2009}. The optimal weighting
scheme proposed by \citet{Hamausetal2010} is based on the idea of
shot-noise minimization in the measurements of halo density
fluctuations. The complex study of \citet{Caietal2011} derives, for
various scenarios and among other things, an optimal weight for
estimating the matter density fluctuations from measurements of the
halo density fluctuations. These last two weighting schemes are
complicated to implement in our calculations here. However, in the
limit where the haloes included in the estimates have sufficiently low
masses (i.e. mass-detection threshold for haloes is low), both latter
weighting schemes converge towards mass weighting.

\noindent The standard FKP weighting is:
\be w_{\rm FKP}(\bx(\chi,{\bm\Omega}),M) = w_{\rm FKP}(\chi)
= \frac{1}{(1+\alpha)+\nbarhx P_i} \ .\ee
We shall denote the mass weighting of the clusters
\citep{Seljaketal2009,Hamausetal2010} combined with FKP's space
weighting as,
\be w_{\rm M+FKP}(\bx(\chi,{\bm\Omega}),M) = 
w_{\rm M+FKP}(\chi,M) = M w_{\rm FKP}(\chi)
 \ .\ee
Before proceeding further, it will be very useful to introduce the
following quantities:
\ba
\nbarhx & \equiv & \int_{0}^{\infty} dM \nbar(M,\chi) \Theta(\chi | M) 
= \int_{\Mlimx}^{\infty}dM \nbar(M,\chi)\ ; \nn \\
\rhobhx & \equiv &  \int_{0}^{\infty} dM M  \nbar(M,\chi) \Theta(\chi | M) 
= \int_{\Mlimx}^{\infty}dM M \nbar(M,\chi) \ ; \nn \\
\overline{b^j}(\chi) & \equiv & \frac{D^j(\chi)}{\nbarhx} 
\int_{0}^{\infty} dM \nbar(M,\chi) b^j(M,\chi) \Theta(\chi | M)
= \frac{D^j(\chi)}{\nbarhx} \int_{\Mlimx}^{\infty} \hspace{-0.2cm} dM \nbar(M,\chi) b^j(M,\chi) \ ; 
\hspace{0.2cm} \mbox{and}\hspace{0.2cm} \bbar(\chi) \equiv \overline{b^1}(\chi) \ ; \nn \\
\overline{b_M^j}(\chi) & \equiv & \frac{D^j(\chi)}{\rhobhx} \int_0^{\infty} dM M \nbar(M,\chi)
 b^j(M,\chi) \Theta(\chi | M) = \frac{D^j(\chi)}{\rhobhx}\int_{\Mlimx}^{\infty}\hspace{-0.7cm} 
dM M \nbar(M,\chi) b^j(M, \chi)\ ; \hspace{0.1cm} \mbox{and}\hspace{0.1cm} \overline{b_M}(\chi) 
\equiv \overline{b_M^1}(\chi) \ ; \nn \\
\left<M^j(\chi)\right> & \equiv & \frac{1}{\nbarhx} \int_0^{\infty} dM M^j \nbar(M,\chi) \Theta(\chi | M)
=  \frac{1}{\nbarhx}\int_{\Mlimx}^{\infty}dM M^j \nbar(M,\chi) \ ; 
\hspace{0.2cm} \mbox{and}\hspace{0.2cm} \left<M(\chi)\right> \equiv \left<M^1(\chi)\right> \ .
\label{eq:quants}
\ea

\begin{figure}
\centering{
  \includegraphics[width=10cm,clip=]{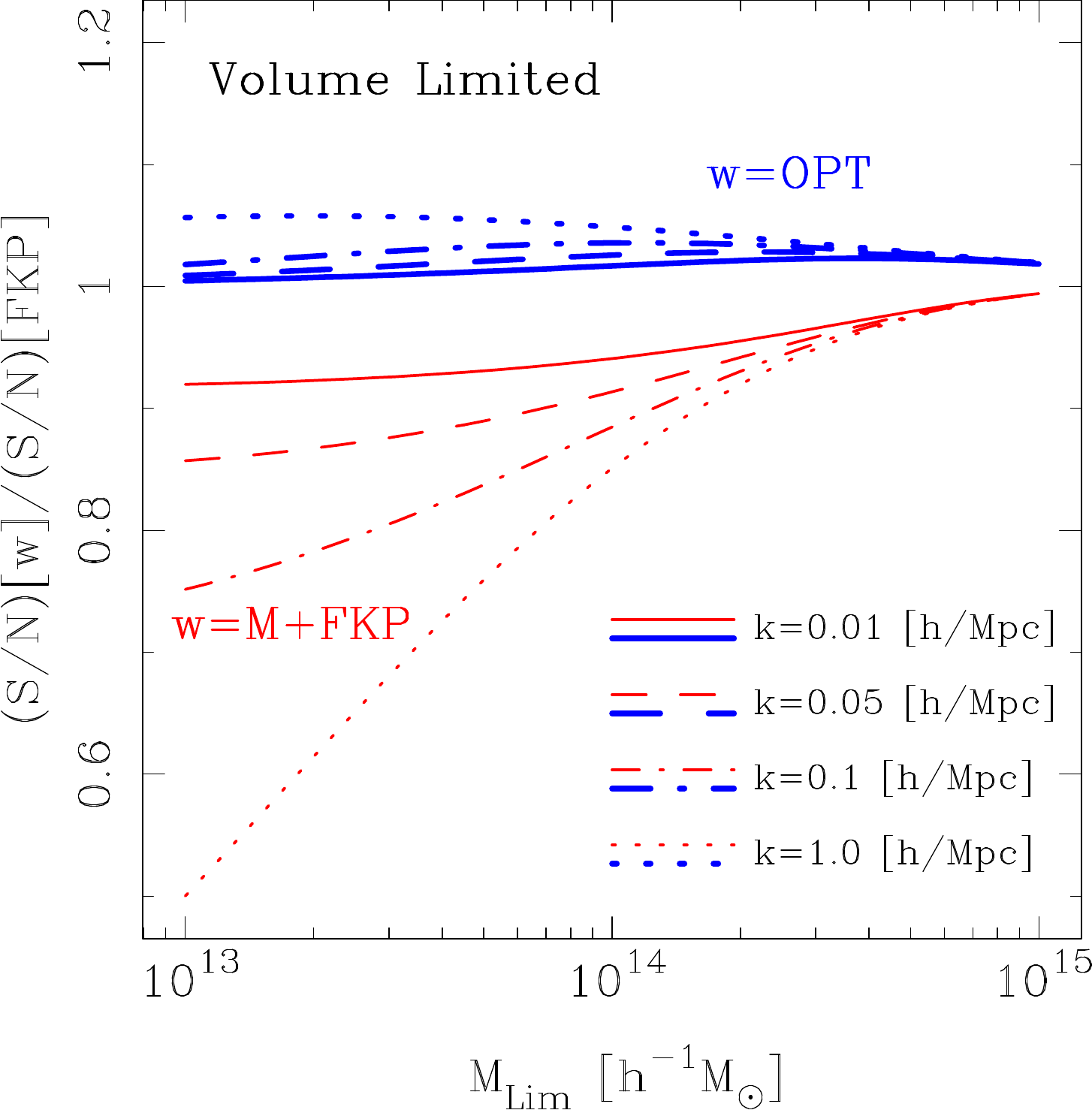}}
\caption{\small{Signal-to-noise ratios for the optimal and mass
    weighted cluster power spectrum relative to the FKP
    signal-to-noise, as a function of the minimum detectable mass in a
    volume-limited cluster survey. Thick blue lines and thin red lines
    denote the optimal and mass-weighting schemes, respectively. The
    solid, dash, dot-dash and dotted lines show the results for
    band-power wavenumbers $k= \{0.01,\, 0.05,\, 0.1,\,1.0\}
    \kMpc$. }\label{fig:SNVolLim}}
\end{figure}

We now proceed to the calculation of the $\SN$ for the matter power
spectrum using the optimal weights given by \Eqn{eq:optimalweight}, as
well as $w_{\rm FKP}$ and $w_{\rm M+FKP}$. Using \Eqn{eq:Func1}, one
can show the following general expressions:
\ba
\SNpar^2_{\rm FKP}\hspace{-0.5cm}(k_i) & = &\frac{\Omu V_i}{2(2\pi)^3} 
\left\{\int_0^{\infty} d\chi \,\chi^2 \left[\frac{\bbar(\chi)\nbarhx \overline{P}_i}
{(1+\alpha)+\nbarhx \overline{P}_i}\right]^2 \right\}^2\Bigg/
\int_0^{\infty} d\chi \,\chi^2 \left\{ \nbarh(\chi) \overline{P}_i
\frac{\left[(1+\alpha)+\bbar^2(\chi) \nbarhx \overline{P}_i\right]}
{\left[(1+\alpha)+\nbarhx \overline{P}_i\right]^2} \right\}^2 \ ; 
\label{eq:SNfkp} \\
\SNpar^2_{\rm OPT}\hspace{-0.5cm}(k_i) & = &\frac{\Omu V_i}{2(2\pi)^3}
\int_0^{\infty} d\chi \,\chi^2 \left[\frac{\nbarhx \,\bbarsqx \, \overline{P}_i}
{(1+\alpha)+\nbarhx \, \bbarsqx \, \overline{P}_i}\right]^2  \ ;
\label{eq:SNopt} \\ 
\SNpar^2_{\rm M+FKP}\hspace{-1cm}(k_i) & = & \frac{\Omu V_i}{2(2\pi)^3}
\left\{\int_0^{\infty} \hspace{-0.2cm} d\chi \,\chi^2\left[\frac{\overline{b_M}(\chi) \rhobhx \overline{P}_i}
{(1+\alpha)+\nbarhx \overline{P}_i}\right]^2\right\}^2 
\hspace{-0.2cm}\Bigg/  \hspace{-0.2cm}
\int_0^{\infty} \hspace{-0.2cm}d\chi \, \chi^2 \left\{ \overline{P}_i  \frac{
(1+\alpha)\left<M^2(\chi)\right> \nbarhx + \left[\overline{b_M}(\chi) \rhobhx\right]^2 
\overline{P}_i}{\left[(1+\alpha)+\nbarhx \overline{P}_i\right]^2}\right\}^2 
\label{eq:SNmass}
\ea
\subsection{Volume-limited samples}
To begin, we adopt the selection function for a volume-limited sample
as described in \S\ref{sec:survey}. We also note that for a
volume-limited sample, the weight function possesses no spatial
dependence and so we are free to take the weights simply as
\be 
w_{\rm FKP}(\bx, M) = 1 \ ; \hspace{0.2cm} w_{\rm OPT}(\bx, M) = b(M) \ ;
\hspace{0.2cm} w_{\rm M+FKP}(\bx, M) = M \ .
\label{eq:weightsVL}
\ee

Let us now consider how the $\SN$ behaves for the three schemes
mentioned above, as a function of the minimum detectable mass
$\Mlim$. Note that in the volume limited sample we shall assume that
$\Mlimx=\Mlim(\chi_0)$ is a constant throughout the survey. With the
weights given by \Eqn{eq:weightsVL}, Eqs.~(\ref{eq:SNfkp}),
~(\ref{eq:SNopt}), ~(\ref{eq:SNmass}) transform into:

\ba
\SNpar_{\rm FKP}\hspace{-0.5cm}(k_i) & = & \sqrt{\frac{\Vu V_i}{2(2\pi)^3}} \left[ 
\frac{\bbar^2 \nbarh \overline{P}_i}{(1+\alpha)+\bbar^2 \nbarh \overline{P}_i}\right] \ . \nn \\
\SNpar_{\rm OPT}\hspace{-0.5cm}(k_i) & = & \sqrt{\frac{\Vu V_i}{2(2\pi)^3}} \left[ 
\frac{\bbarsq \nbarh \overline{P}_i}{(1+\alpha)+\bbarsq \nbarh \overline{P}_i}\right] \ . \nn \\
\SNpar_{\rm M+FKP}\hspace{-1cm}(k_i) & = & \sqrt{\frac{\Vu V_i}{2(2\pi)^3}} \left[ 
\frac{r_M^2\bbarM^2 \nbarh \overline{P}_i}{(1+\alpha)+r_M^2 \bbarM^2 \nbarh \overline{P}_i}\right] \ ,
\label{eq:SN_VL}
\ea
where we defined $r_M^2\equiv \left<M\right>^2/\left<M^2\right>$. All
quantities on the right-hand side of these equations are the same as
defined by \Eqn{eq:quants}, only evaluated for a constant mass
thereshold $\Mlim(\chi_0)=\Mlim$. The $\SN$ values obtained from the
three weighting schemes may be more easily compared if we take the
ratio of the optimal and mass weighting scheme with respect to the
original FKP weights. Whereupon,
\ba
\frac{\left({\mathcal S}/{\mathcal N}\right)_{\rm OPT}}
{\left({\mathcal S}/{\mathcal N}\right)_{\rm FKP}} = \frac
{1+(1+\alpha)/\nbarh \overline{P}_i \bbar^2 }
{1+(1+\alpha)/\nbarh \overline{P}_i \bbarsq }\ ; \hspace{0.5cm}
\frac{\left({\mathcal S}/{\mathcal N}\right)_{\rm M+FKP}}
{\left({\mathcal S}/{\mathcal N}\right)_{\rm FKP}} = 
\frac{1+(1+\alpha)/\nbarh \overline{P}_i \bbar^2}
{1+(1+\alpha)/\nbarh \overline{P}_i \bbarM^2 r_M^2  } \ .
\label{eq:FbFu} 
\ea
There are two limiting cases of interest: 
\begin{itemize}
\item $\nbarh \overline{P}_i\gg 1$: in this limit the two terms in
  \Eqn{eq:FbFu} become,
\ba
\frac{\left({\mathcal S}/{\mathcal N}\right)_{\rm OPT}}
{\left({\mathcal S}/{\mathcal N}\right)_{\rm FKP}} \approx 
1+\frac{(1+\alpha)}{\nbarh P_i}\left[\frac{1}{\bbar^2}-
\frac{1}{\bbarsq} \right]  \ge 1 \  \ ; \hspace{0.5cm}
\frac{\left({\mathcal S}/{\mathcal N}\right)_{\rm M+FKP}}
{\left({\mathcal S}/{\mathcal N}\right)_{\rm FKP}} \approx 
1+\frac{(1+\alpha)}{\left(\nbarh P_i\right)}
\left[\frac{1}{\bbar^2}-\frac{1}{\bbarM^2r_M^2} \right]\  \ . 
\label{eq:ineqq1}
\ea
\item $\nbarh \overline{P}_i\ll 1$: in this limit the expressions in
  \Eqn{eq:FbFu} become,
\ba
\frac{\left({\mathcal S}/{\mathcal N}\right)_{\rm OPT}}
{\left({\mathcal S}/{\mathcal N}\right)_{\rm FKP}} \approx 
\frac{ \bbarsq}{\bbar^2} \ge 1 \  \ ; \hspace{0.5cm}
\frac{\left({\mathcal S}/{\mathcal N}\right)_{\rm M+FKP}}
{\left({\mathcal S}/{\mathcal N}\right)_{\rm FKP}} \approx 
 \frac{1}{r_M^2}\left(\frac{\bbarM}{\bbar}\right)^2  \ .
\label{eq:ineqq2} 
\ea
\end{itemize}
The first inequalities given by \Eqns{eq:ineqq1}{eq:ineqq2} both
follow from the fact that $\bbar^2\le \bbarsq \label{eq:ineq1}$ (for a
proof of this relation see Appendix~\ref{app:ineq1}).  In order to
determine whether the second expressions in
\Eqns{eq:ineqq1}{eq:ineqq2} are greater or less than unity, one must
examine the product of the quantities ${\bbar^2}/{\bbarM^2}\le1$ and
${\left<M^2\right>}/{ \left<M\right>^2}\ge1$. Unfortunately, this is
not so easy to determine and we therefore turn to numerical evaluation
of the expressions.

In \Fig{fig:SNVolLim} we present the evolution of the $\SN$ for the
optimal and mass weighting schemes, relative to the $\SN$ obtained
from the FKP weighting, as a function of $\Mlim$. The blue and red
lines denote the optimal and mass-weighting schemes, respectively and
results are presented for several $k$-band powers.  Here we have used
the LCDM model as a particular example and have adopted the bias and
dark matter halo mass function formula of \citet{ShethTormen1999} to
evaluate \Eqn{eq:FbFu}. We notice the following: the optimal weighting
scheme does indeed maximize the $\SN$; the mass-weighting scheme is
inferior to the optimal and FKP weighting schemes; the overall gains
of the optimal weighting scheme relative to the FKP scheme appear
modest $\sim5\%$.
\subsection{Flux-limited samples}
\label{ssec:fluxlim}
For flux-limited samples, the $\SN$ for the three weighting schemes is
given by the general Eqs.~(\ref{eq:SNfkp})-(\ref{eq:SNmass}).  Further
analytic developments are non-trivial, and so we numerically evaluate
them for the particular case of LCDM. To accomplish this we require
knowledge of the minimum detectable mass as a function of $\chi$. As
mentioned in \S\ref{sec:survey}, $\Mlimx$, in general, is a
complicated function of the survey flux-limit and the cluster
identification algorithm. For simplicity, we shall assume that this
can be written as:
\be \Mlim(\chi) = \Mlim(0) \ex^{\gamma z(\chi)} \label{eq:min} \ .\ee
If we adopt the value $\gamma=1$, the above functional form roughly
matches the cluster selection as a function of redshift, which one
finds for weak lensing detected cluster surveys
\citep{MarianBernstein2006}. We evaluate the above $\SN$ ratios as a
function of $\Mlim(0)$, with $\Mlim(0)$ varying in the range
$\left[10^{13},10^{15}\right]\Msol$.

\begin{figure}
\centering{
  \includegraphics[width=10cm,clip=]{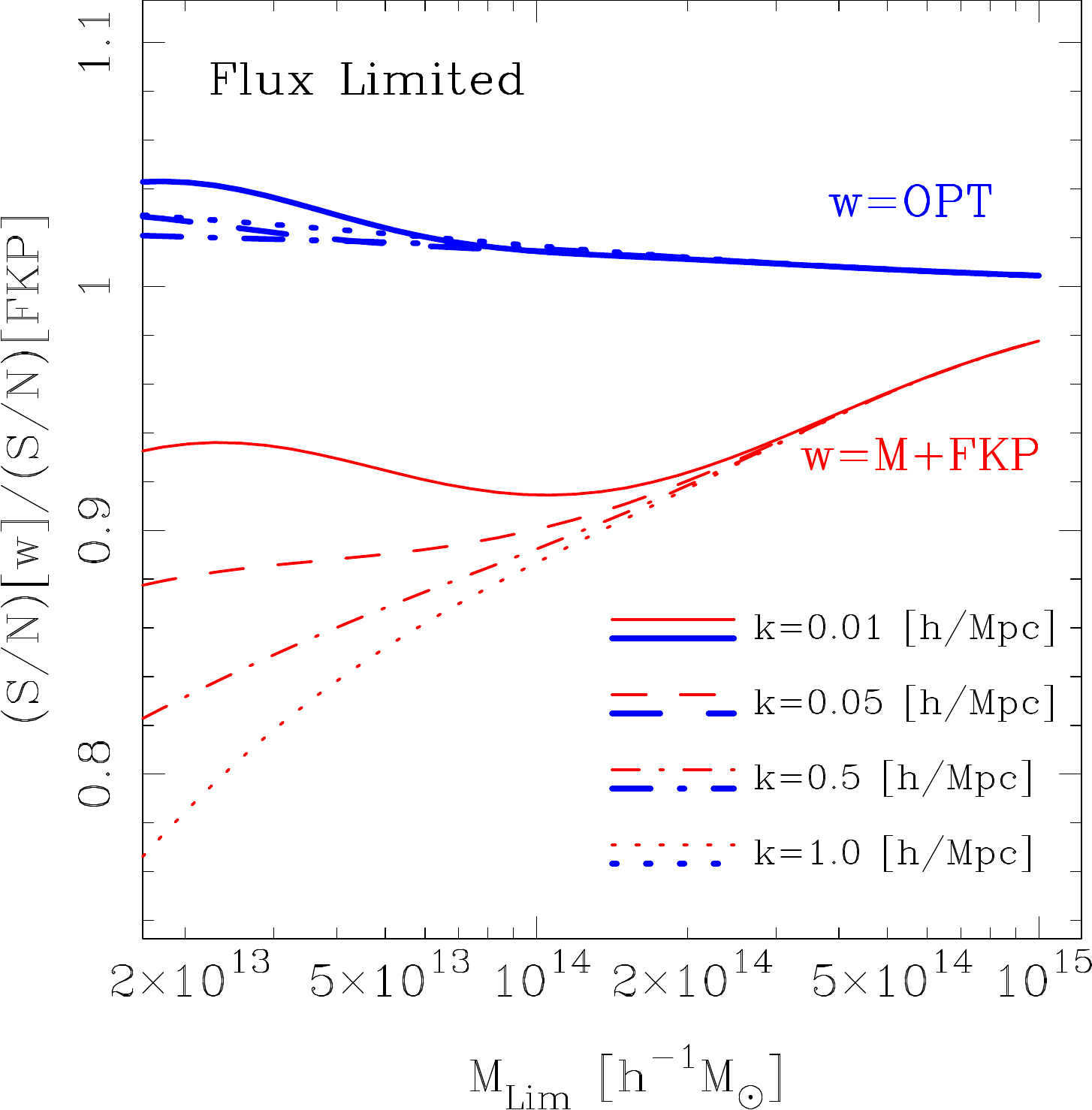}}
\caption{\small{Same as \Fig{fig:SNVolLim} except this time for a
    flux-limited survey.\label{fig:SNFluxLim}}}
\end{figure}

\Fig{fig:SNFluxLim} is the analogue of \Fig{fig:SNVolLim} for a
flux-limited cluster survey. The blue and red lines denote the optimal
and mass weighting schemes, respectively and results are presented for
several $k$-band powers.  Again, we have used the LCDM model as an
example and have adopted the bias and dark matter halo mass functions
from \citet{ShethTormen1999} to evaluate Eqns~(\ref{eq:SNfkp}),
(\ref{eq:SNopt}) and (\ref{eq:SNmass}). Several features may be noted:
the optimal weighting scheme always maximizes the signal-to-noise
ratio; the mass-weighting scheme is inferior to both the optimal and
FKP schemes for all scales; the overall $\SN$ gains for the optimal
weighting are very modestly ($\sim3\%$) improved over the FKP
approach.

\begin{figure}
\centering{
  \includegraphics[width=10cm,clip=]{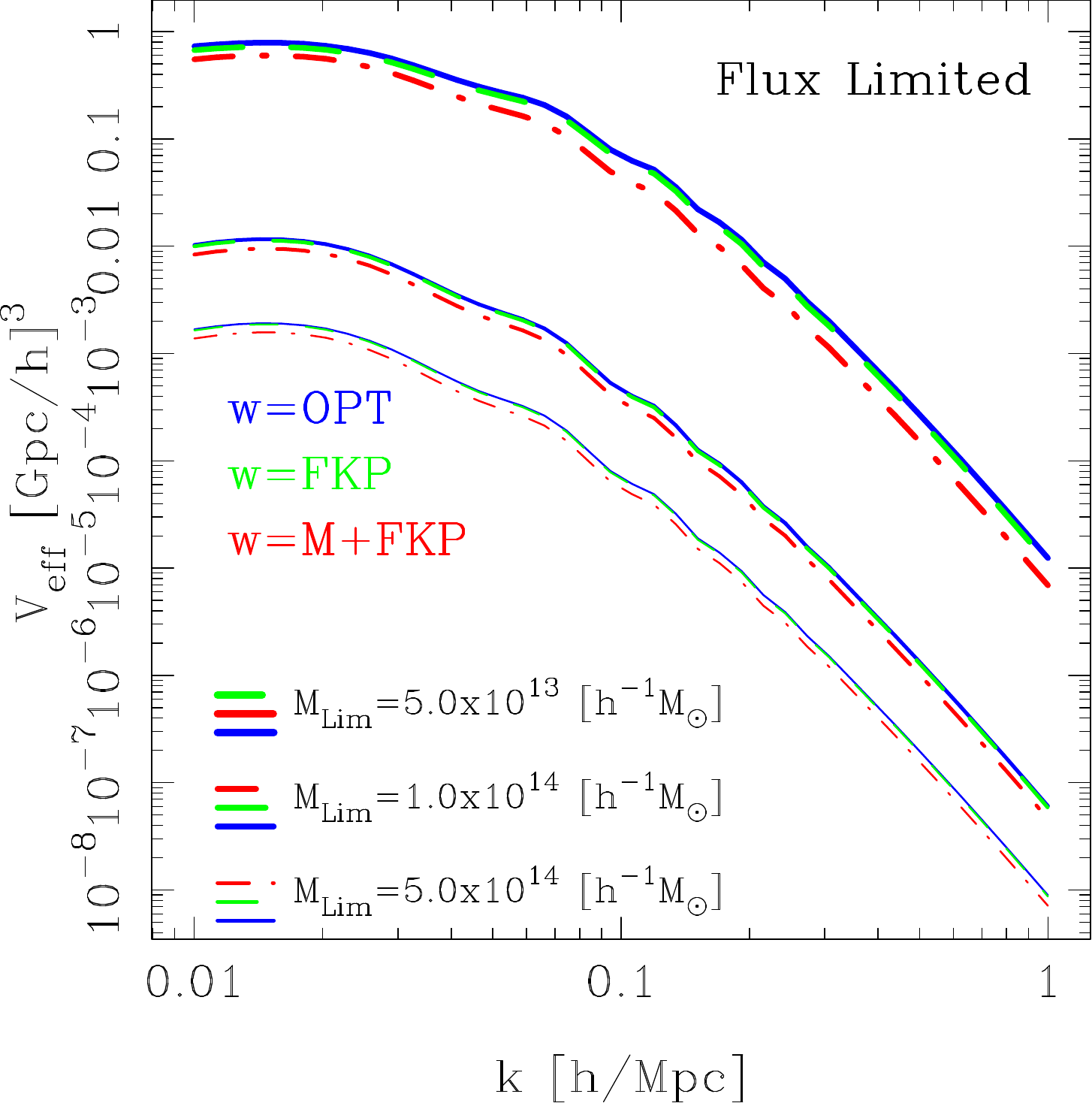}}
\caption{\small{Effective survey volume as a function of the power
    spectrum wavenumber probed. Lines of decreasing thickness
    correspond to a minimum detectable mass $\Mlim(0)$, of
    $5.0\times10^{13}\Msol$, $1.0\times10^{14}\Msol$,
    $5.0\times10^{14}\Msol$, respectively. The solid, dashed and
    dot-dashed lines correspond to $V_{\rm eff}$ for optimal, FKP and
    M+FKP weighting, respectively.}\label{fig:Veff}}
\end{figure}
\subsection{The Fisher matrix}
As a final corollary to this section, we explore the cosmological
information content of an optimally weighted cluster power spectrum
analysis. Following \citet{Tegmark1997}, the Fisher information matrix
for a power spectrum analysis can be defined as:
\be 
{\mathcal F}_{\alpha\beta}  = \sum_{i,j} 
\frac{\partial \overline{P}_i}{\partial \alpha}C^{-1}_{ij}
\frac{\partial \overline{P}_j}{\partial \beta}
 =  \sum_{i,j}\frac{\partial \log \overline{P}_i}{\partial
    \alpha} \overline{P}_i C^{-1}_{ij}\overline{P}_j 
\frac{\partial \log \overline{P}_j}{\partial \beta}\ ,
\ee
where $\partial/\partial\alpha \equiv
\partial/\partial\theta_{\alpha}$ denote partial derivatives with
respect to the cosmological parameters $\theta_{\alpha}$, and where
here we have neglected the information content in the covariance
matrix \citep{Tegmarketal1997}.  On inserting our earlier expression
for the covariance matrix, as given by \Eqn{eq:CovGauss5}, into the
above expression, then the Fisher matrix becomes:
\be 
{\mathcal F}_{\alpha\beta}  =  \sum_{i,j}\frac{\partial \log
  \overline{P}_i}{\partial \alpha} \overline{P}_i
\frac{\delta^{K}_{ij}}{\sigma^2_P(k_i)} \overline{P}_j \frac{\partial
  \log \overline{P}_j}{\partial \beta} = 
\sum_{i}\frac{\partial \log \overline{P}_i}{\partial \alpha}
\frac{\partial \log \overline{P}_i}{\partial \beta}
\left(\frac{{\mathcal S}}{{\mathcal N}}\right)_{\rm OPT}^2\!\!(k_i)\ .  
\ee
On inserting our expression for the $\SN$ for the optimal weights
given by \Eqn{eq:SNopt}, the Fisher matrix, in the continuum limit of
Fourier modes, can be written:
\be {\mathcal F}_{\alpha\beta} =  \frac{1}{2}\int \frac{\dk}{(2\pi)^3}
\frac{\partial \log P(k)}{\partial \alpha}
\frac{\partial \log P(k)}{\partial \beta} V_{\rm eff}(k)\ , \label{eq:Fisher}
\ee
where the effective survey volume has the new form:
\be V_{\rm eff}(k)=\int_0^{\infty} d\chi \chi^2 \left[\frac{\nbarhx\bbarsqx P(k)}
{(1+\alpha)+\nbarhx\bbarsqx P(k)}\right]^2 \ .\ee

In \Fig{fig:Veff} we show the effective survey volume as a function of
wavenumber, for a flux-limited survey of similar type to that
described in \S\ref{ssec:fluxlim}. Here we consider the three cases
where the minimum detectable mass normalization parameter from,
c.f. \Eqn{eq:min}, has the values $5.0\times10^{13}\Msol$,
$1.0\times10^{14}\Msol$, $5.0\times10^{14}\Msol$, respectively. In all
cases the optimal weighting increases $V_{\rm eff}$. Thus we conclude
that the cosmological information extractable from an optimal weighted
cluster survey will exceed that from the sub-optimal strategies, such
as FKP or M+FKP weighting. \Eqn{eq:Fisher} may thus be used as the
starting point for exploring the cosmological information content of
optimally weighted cluster power spectra \citep[for a review of the
  Fisher matrix approach, see][]{Heavens2009}.

\section{Conclusions}\label{sec:conclusions}

In this paper we have developed a formalism to compute weights
maximizing the $\SN$ when estimating the cluster power spectrum, and
used it to make inferences about the information in the matter power
spectrum. Our derivation generalizes the original approach of FKP for
galaxies, and is analogous to the derivation by PVP for examining the
impact of luminosity dependent galaxy bias on the weights. Our
derivation provides for the first time a completely analytic proof for
the optimal weight equation, and it also corrects several errors that
were found in these earlier works.

In \S\ref{sec:survey} we described the generic properties of a
cluster survey, taking into account finite survey geometry, arbitrary
weighting of position and mass, and a flux limit; we also introduced
out statistical treatment for describing the cluster density field
using delta function expansions.

In \S\ref{sec:twopoint} we presented estimators for recovering the
matter clustering from the computation of the two-point correlation
function and the power spectrum of the cluster field. We demonstrated
that in order to extract information from the matter clustering one
must deconvolve the survey window function. For large homogeneous
survey volumes, provided an appropriate choice for the normalization
of the cluster field is taken, this estimate was shown to be an
unbiased estimator of the dark matter power spectrum, modulo a
shot-noise correction.

In \S\ref{sec:fourpoint} we explored the statistical fluctuations in
the power spectrum of clusters. We derived general expressions for the
covariance of the cluster sample, including all non-Gaussian terms
arising from the nonlinear evolution of matter fluctuations and
discreteness effects. This generalized the result of
\citet{Smith2009}. We then proved the necessary conditions for the
covariance matrix to be diagonal. 
 
In \S\ref{sec:optimal} we have provided an analytic derivation of the
optimal weights for a cluster power spectrum analysis. We show in
general terms that the optimal weights are separable functions of mass
and space. 

In \S\ref{sec:case} we presented a comparison of the optimal weighting
scheme with the original FKP scheme and with a mass-weighting scheme.
The latter was advocated by \citet{Seljaketal2009}, later as a linear
function of mass by \citet{Hamausetal2010}, in the context of reducing
stochasticity in halo fields. In the limit of a low-mass cluster
detection threshold the study of \citet{Caietal2011} also found mass
weighting to be optimal for matter density field reconstruction. For
the case of both volume- and flux-limited cluster surveys the optimal
weighting scheme outperforms both alternate weighting schemes. The
gains over the FKP method are very modest $\sim5\%$. Mass weighting
performs significantly worse than the optimal scheme, with a relative
loss in $\SN$ of $\sim10\%$ on large scales, and of $\sim20\%$ on
intermediate scales.  Whilst the mass-dependent weighting may be
useful for reconstructing the matter field from a cluster sample, we
recommend that it should not be used to extract cosmological
information from cluster power spectrum analysis. We also presented a
new expression for the Fisher information matrix, for an
optimally-weighted cluster power spectrum measurement. 

In this paper we have derived the optimal weights for measuring the
cluster power spectrum under certain conditions, if these conditions
are relaxed then the weights are no longer optimal. It will be
interesting in future work to explore whether a more general weighting
scheme can be derived for the more realistic situations where a
non-diagonal covariance matrix is considered. We also expect that the
weights that we have derived for the power spectrum should also be
used to obtain optimal measurements of the cluster correlation
function. However, we have not yet demonstrated this explicitly.


\section*{Acknowledgements}

We thank Ravi Sheth and Simon White for useful discussions.  RES
acknowledges support from ERC Advanced grant 246797 GALFORMOD.  LM
thanks MPA for its kind hospitality while this work was being
performed. LM was partly supported by the DFG through the grant MA
4967/1-2.



\bibliographystyle{mnras}
\bibliography{/Users/res/WORK/PAPERS/REFS/refs.bib}


\appendix
\section{Computing the 2-point correlations}
\label{app:AI}
\noindent Throughout this section we shall use the shorthand notation
$\langle n_\c n'_\c\rangle \equiv \langle n_\c(\bx, M) n_\c(\bx', M')
\rangle$, and similar for $\langle n_\c n'_s\rangle,\, \langle n_s
n'_s \rangle$.  The evaluation of these terms follows closely that of
$<\Fc>$. Proceeding with the first term, we write:
\be
\left<n_\c n'_\c \right>  = \int \prod_{l=1}^{N} \left\{\dx_l \dM_l\right\}
p(\bx_1,\dots,\bx_N,M_1,\dots,M_N) \sum_{i,j}^{N}\delta^{\rm D}(\bx-\bx_i)
\delta^{\rm D}(M-M_i)\delta^{\rm D}(\bx'-\bx_j) \delta^{\rm D}(M'-M_j)  \ .
\nn \ee
The double sum can be broken up into the terms where $(i\ne j)$ and
the terms where $(i=j)$, whereupon
\ba
\left<n_\c n'_\c\right> & = & \int \prod_{l=1}^{N} \left\{\dx_l \dM_l\right\} 
p(\bx_1,\dots,\bx_N,M_1,\dots,M_N) \left[\sum_{i\ne j}^{N} 
\delta^{\rm D}(\bx-\bx_i)\delta^{\rm D}(M-M_i)\delta^{\rm D}(\bx'-\bx_j)
\delta^{\rm D}(M'-M_j)\right. \nn \\ 
& & + \left.\sum_{i=j}^{N} \delta^{\rm D}(\bx-\bx_i)\delta^{\rm D}(\bx'-\bx_i)
\delta^{\rm D}(M-M_i)\delta^{\rm D}(M'-M_i) \right] \nn \\
& = & 
\sum_{i\ne j}^{N} p(\bx,\bx', M, M') + \sum_{i=j}^{N} p(\bx, M) 
\delta^{\rm D}(\bx-\bx')\delta^{\rm D}(M-M') \label{eq:ncnc0} \ .
\ea
In order to proceed further, we need to specify the joint probability
density distribution for obtaining clusters at positions $\bx$ and
$\bx'$ and with masses $M$ and $M'$. This we do through the introduction 
of correlation functions:
\be p(\bx, \bx', M, M')\equiv 
p(\bx, M)p(\bx', M')\left[1+\xi^{\c}(\bx, \bx', M, M')\right] 
= \frac{\nbar(M) \nbar(M')}{N^2} \left[1+\xi^{\c}(\bx, \bx', M, M')\right] \ ,
\nn \ee
where $\xi^{\rm c}(\bx, \bx', M, M')$ is the two-point
cross-correlation function of clusters with masses $M$ and
$M'$. On use of the above definition in \Eqn{eq:ncnc0}, we find
\be
\left<n_\c n'_\c \right> \approx  \frac{}{}\nbar(M)\nbar(M')
\left[1+\xi^\c(\bx, \bx', M, M')\right]+\nbar(M)\delta^{\rm D}(\bx-\bx')
\delta^{\rm D}(M-M')  \ ,
\label{eq:ncnc}
\ee
where in arriving at the last equality we have assumed that $N\gg 1$
and therefore $N-1\approx N$.  We may now write down directly the
remaining expectation values that enter \Eqn{eq:FF}, whereupon:
\ba 
\left<n_\c n'_s\right> & = & \alpha^{-1}\nbar(M) \nbar(M') \ ; \nn \\
\langle n_s n'_s \rangle & = & \alpha^{-2} \nbar(M) \nbar( M')
+\alpha^{-1}\nbar(M)\delta^{\rm D}(\bx-\bx')\delta^{\rm D}(M-M') \ .
\label{eq:ncns}
\ea
%
\section{Statistical fluctuations in the $\Fc$ power spectrum}
\label{app:AII}
\subsection{General expression of the covariance of the power spectrum estimator}
\label{app:AII.1}
The covariance of the matter power spectrum estimator given by \Eqn{eq:Pest}
is formed of four terms:
\be 
{\rm Cov}\!\left[\hat{P}(\bk_1),\hat{P}(\bk_2)\right] = 
{\rm Cov}\!\left[|\Fcf(\bk_1)|^2,|\Fcf(\bk_2)|^2\right] 
+{\rm Var}\!\left[P_{\rm shot}\right]
-{\rm Cov}\!\left[|\Fcf(\bk_1)|^2, P_{\rm shot}\right] 
-{\rm Cov}\!\left[|\Fcf(\bk_2)|^2, P_{\rm shot}\right] \ ,
\label{eq:CovPfull}
\ee
where 
\ba 
{\rm Cov}\!\left[|\Fcf(\bk_1)|^2,|\Fcf(\bk_2)|^2\right] & \equiv & 
\left<|\Fcf(\bk_1)|^2|\Fcf(\bk_2)|^2\right>-
\left<|\Fcf(\bk_1)|^2\right> \left<|\Fcf(\bk_2)|^2 \right> \ ;\nn \\
{\rm Var}\!\left[P_{\rm shot}\right] & \equiv & 
\left<[P_{\rm shot}]^2\right>-\left<P_{\rm shot}\right>^2 \ ;\nn\\
{\rm Cov}\!\left[|\Fcf(\bk_i)|^2, P_{\rm shot}\right] & \equiv & 
\left<|\Fcf(\bk_i)|^2 P_{\rm shot}\right>-
\left<|\Fcf(\bk_i)|^2\right> \left<P_{\rm shot}\right> \ \ \ ; \ \ i \in (1,2) \ .
\ea
Assuming that the total number density of clusters is a deterministic
quantity, all of the covariance terms involving $P_{\rm shot}$
vanish. Thus we are left with the task of determining the covariance
of the modulus square of the $\Fcf$ field, i.e. \Eqn{eq:CovPest1}.
\subsection{Derivation of the covariance of the power spectrum estimator}
\label{app:AII.2}
For a succint presentation of the covariance calculation, we make use
of the following shorthand notation:
\be 
\nbar_i\equiv \nbar(M_i) \:;\:\: w_i\equiv w(\bx_i, M_i) \: ; \:\:
b_i\equiv b(M_i) \: ;\:\: \Theta_i\equiv \Theta(\bx_i|M_i) \: ; \:\:
n_{\c, i} \equiv n_\c(\bx_i, M_i) \: ;\:\:
n_{s, i} \equiv n_s(\bx_i, M_i) \: ;\:\:
\delta^{\rm D}_{ij} \equiv \delta^{\rm D}(\bx_i-\bx_j) \delta^{\rm D}(M_i-M_j)  \ .
\label{eq:shorts}
\ee
The covariance matrix of the power in the field $\Fc$ can be written:
\be
{\rm Cov}\!\left[|\Fcf(\bk_1)|^2,|\Fcf(\bk_2)|^2\right] 
=
\int \dk_3 \,\dk_4 \delta^{\rm D}(\bk_1+\bk_3) \delta^{\rm D}(\bk_2+\bk_4)
\left[
\left<\Fcf(\bk_1)\dots \Fcf(\bk_4)\right> \hspace{-0.05cm}-\hspace{-0.05cm}
\left<\Fcf(\bk_1)\Fcf(\bk_3)\right>\hspace{-0.08cm}\left<\Fcf(\bk_2)\Fcf(\bk_4)\right> \right]  \hspace{-0.1cm}
\label{eq:one}
\ee 
In order to proceed further we see that we must compute the four-point
correlation function of the Fourier modes $\Fcf(\bk)$. This is
equivalent to specifying the four-point correlation function of the
field $\Fc$:
\be 
\left<\Fcf(\bk_1)\dots \Fcf(\bk_4)\right> = \int \dx_1 \dots \dx_4 
\left<\Fc(\bx_1)\dots \Fc(\bx_4)\right>
\ex^{i\bk_1\cdot\bx_1+\dots+\bk_4\cdot\bx_4} \ .
\ee
Substituting \Eqn{eq:clusterden} into the expression for the
four-point correlation function, we find:
\be \left<\Fc(\bx_1)\dots \Fc(\bx_4)\right>  = 
\frac{1}{A^2} \int \prod_{i=1}^4 \left\{\dM_i w(\bx_i,M_i)\Theta(\bx_i|M_i)\right\}
\left<\frac{}{}\left[n_\c(\bx_1,M_1)-\alpha n_s(\bx_1,M_1)\right]
\dots\left[n_\c(\bx_4,M_4)-\alpha n_s(\bx_4)\right]\right>  \ .
\label{eq:FFFF}
\ee
Expanding the term in angled braces on the right-hand side gives,
\ba  \left<\left[n_{\c,1}-\alpha n_{s,1}\right]
      \dots\left[n_{\c,4}-\alpha n_{s,4}\right]
\right> & = & \left<n_{\c,1}\dots n_{\c,4}\right>  -\alpha
\left[\frac{}{}\left<n_{\c,1}n_{\c,2}n_{\c,3}n_{s,4}\right> 
+{3\rm cyc}\right] +\alpha^2
\left[\frac{}{}
  \left<n_{\c,1}n_{\c,2}n_{s,3}n_{s,4}\right>+{5\rm
    perm}\right] \nn \\ & & -\alpha^3 \left[\frac{}{}
  \left<n_{\c,1}n_{s,2}n_{s,3}n_{s,4}\right> +{3\rm cyc}\right] 
+\frac{}{}\alpha^4\left<n_{s,1}\dots n_{s,4}\right> \ .
\label{eq:expec1}
\ea
We focus on the first term in curly brackets on the right-hand side,
and insert our delta function expansion for the number density field:
\ba
\left<n_{\c, 1}\dots n_{\c, 4}\right> & = &
\left<\sum_{i_1,i_2, i_3, i_4 = 1}^{N}\delta^{\rm D}(\bx_1-\bx'_{i_1})\dots 
\delta^{\rm D}(\bx_4-\bx'_{i_4})\delta^{\rm D}(M_1-M'_{i_1})\dots 
\delta^{\rm D}(M_4-M'_{i_4})\right>\nn \\
 & = & 
\int \prod_{p=1}^{N} \left\{\dx'_p \,\dM'_p\right\} p(\bx'_1,\dots,\bx'_N, M'_1,\dots, M'_N)
\left\{\Gamma_1+\Gamma_2+\Gamma_3+\Gamma_4+\Gamma_5\right\} \ ,
\label{eq:L}
\ea
where in the above we have defined the following terms:
\ba 
\Gamma_1 & \equiv & 
\sum_{i_1\ne i_2\ne i_3\ne i_4}\delta^{\rm D}(\bx_1-\bx'_{i_1})\dots \delta^{\rm D}(\bx_4-\bx'_{i_4})
\delta^{\rm D}(M_1-M'_{i_1})\dots \delta^{\rm D}(M_4-M'_{i_4}) \ ; \nn \\ 
\Gamma_2 & \equiv & \sum_{i_1 \ne i_2\ne i_3=i_4}
\delta^{\rm D}(\bx_1-\bx'_{i_1})\delta^{\rm D}(\bx_2-\bx'_{i_2})
\delta^{\rm D}(M_1-M'_{i_1})\delta^{\rm D}(M_2-M_{i_2}) \prod_{p=3}^4\left\{
\delta^{\rm D}(\bx_p-\bx'_{i_3})\delta^{\rm D}(M_p-M'_{i_3})\right\} +5 {\rm perms} \ ;
\nn \\
\Gamma_3 & \equiv & 
\sum_{i_1=i_2\ne i_3=i_4}
\prod_{p=1}^2\left\{\delta^{\rm D}(\bx_p-\bx'_{i_1})\delta^{\rm D}(M_p-M'_{i_1})\right\}
\prod_{q=3}^4\left\{\delta^{\rm D}(\bx_q-\bx'_{i_3}\delta^{\rm D}(M_q-M'_{i_3})\right\}
+2 {\rm perms} \ ; \nn \\
\Gamma_4 & \equiv & 
\sum_{i_1\ne i_2=i_3=i_4} \delta^{\rm D}(\bx_1-\bx'_{i_1})\delta^{\rm D}(M_1-M'_{i_1})
\prod_{p=2}^{4}\left\{\delta^{\rm D}(\bx_p-\bx'_{i_2})\delta^{\rm D}(M_p-M'_{i_2})\right\}
 +3 {\rm perms} \ ; \nn \ea
\ba
\Gamma_5 & \equiv & 
\sum_{i_1=i_2=i_3=i_4}\prod_{p=1}^{4}\left\{\delta^{\rm D}(\bx_p-\bx'_{i_1})
\delta^{\rm D}(M_p-M'_{i_1})\right\} \ .
\ea
We now compute the contributions from each of the five $\Gamma$ terms.


\vspace{0.2cm}
\noindent $\bullet$ {\bf Computing $\langle \Gamma_1 \rangle $ : } On
performing the integrations over the delta functions we have:
\ba
\langle \Gamma_1 \rangle & = & \sum_{i_1\ne i_2\ne i_3\ne i_4} 
p(\bx_1,\dots,\bx_4, M_1,\dots,M_4) \nn\\
& = & N(N-1)(N-2)(N-3) p(\bx_1, M_1)\dots p(\bx_4, M_4) 
\left[\frac{}{}\hspace{-0.1cm} 1+\xi^\c_{12}+\xi^\c_{13}+\xi^\c_{14}+\xi^\c_{23}+\xi^\c_{24}+\xi^\c_{34} \right. \nn \\
& & \left. \frac{}{}+\zeta^\c_{123}+\zeta^\c_{234}+\zeta^\c_{341}+\zeta^\c_{412} + 
\xi^\c_{12}\xi^\c_{34}+\xi^\c_{13}\xi^\c_{24}+\xi^\c_{14}\xi^\c_{23}+\eta^\c_{1234}
\right] \nn \\
& \approx & \hspace{-0.2cm}\prod_{p=1}^{4}\left\{\nbar_p\right\}\hspace{-0.1cm}\left[\frac{}{}\hspace{-0.12cm}
1+\xi^\c_{12}+\xi^\c_{13}+\xi^\c_{14} +\xi^\c_{23}+\xi^\c_{24}+\xi^\c_{34}+\zeta^\c_{123}
+\zeta^\c_{234}+\zeta^\c_{341}+\zeta^\c_{412} +\xi^\c_{12}\xi^\c_{34}+\xi^\c_{13}\xi^\c_{24}
+\xi^\c_{14}\xi^\c_{23}+ \eta^\c_{1234} \right], \hspace{1cm}
\label{eq:gamma1}\ea
where the last equality holds in the limit that $N\gg 1$.  The terms
$\zeta^\c$ and $\eta^\c$ are the three- and four-point correlation
functions of the haloes, respectively, and in the above we have also
made use of the short-hand notation:
\be 
\xi^\c_{ij}\equiv\xi^\c(\br_i,\br_j,M_i,M_j) \hspace{0.3cm} ; \hspace{0.3cm}
\zeta^\c_{ijk}\equiv\zeta(\br_i,\br_j,\br_k,M_i,M_j,M_k) \hspace{0.3cm} ; \hspace{0.3cm}
\eta^\c_{ijkl}\equiv \eta^\c(\br_i,\br_j,\br_k,\br_l,M_i,M_j,M_k,M_l) \hspace{0.3cm} .
\ee
%
\vspace{0.2cm}
\noindent $\bullet$ {\bf Computing $\langle \Gamma_2 \rangle $ : } On
integration over the delta functions and using \Eqn{eq:shorts}, we
obtain:
\ba
\langle\Gamma_2\rangle \hspace{-0.2cm} & \equiv & \hspace{-0.6cm}
 \sum_{i_1 \ne i_2 \ne i_3=i_4} \hspace{-0.4cm}  p(\bx_1,\bx_2,\bx_3, M_1, M_2, M_3)\,\delta^{\rm D}_{34}
+\hspace{-0.4cm}\sum_{i_1 \ne i_2=i_3 \ne i_4} \hspace{-0.4cm}  p(\bx_1,\bx_2,\bx_4, M_1, M_2, M_4)\,\delta^{\rm D}_{23} 
+\hspace{-0.4cm}\sum_{i_1=i_2 \ne i_3 \ne i_4} \hspace{-0.4cm}  p(\bx_1,\bx_3,\bx_4, M_1, M_3, M_4)\,\delta^{\rm D}_{12}
\nn \\
& &\hspace{-0.6cm}+ \hspace{-0.4cm}
\sum_{i_1=i_3 \ne i_2 \ne i_4} \hspace{-0.4cm}  p(\bx_1,\bx_2,\bx_4, M_1, M_2, M_4)\,\delta^{\rm D}_{13}
+\hspace{-0.4cm} \sum_{i_1=i_4 \ne i_2\ne i_3} \hspace{-0.4cm}  p(\bx_1,\bx_2,\bx_3, M_1, M_2, M_3)\,\delta^{\rm D}_{14}
+\hspace{-0.4cm} \sum_{i_2=i_4 \ne i_1\ne i_3} \hspace{-0.4cm}  p(\bx_1,\bx_2,\bx_3, M_1, M_2, M_3)\,\delta^{\rm D}_{24}
\ .\nn \ea
On performing the summations and taking the limit of large numbers,
the above expression reduces to:
\ba 
\langle\Gamma_2\rangle \hspace{-0.3cm} & = & \hspace{-0.6cm}
\prod_{p\in\{1,3,4\}}\hspace{-0.4cm}\{\nbar_p\}[1+\xi^\c_{13}+\xi^\c_{14}
+\xi^\c_{34}+\zeta^\c_{134}]\delta^{\rm D}_{12}
+\hspace{-0.4cm}\prod_{p\in\{1,2,4\}}\hspace{-0.4cm}\{\nbar_p\}[1+\xi^\c_{12}+\xi^\c_{24}
+\xi^\c_{24}+\zeta^\c_{124}]\delta^{\rm D}_{23} 
+\hspace{-0.4cm}\prod_{p\in\{1,2,3\}}\hspace{-0.4cm}\{\nbar_p\}
[1+\xi^\c_{12}+\xi^\c_{13}+\xi^\c_{23}+\zeta^\c_{123}]\delta^{\rm D}_{14} \nn \\
& & \hspace{-0.6cm}+\hspace{-0.4cm}
\prod_{p\in\{1,2,4\}}\hspace{-0.4cm}\{\nbar_p\}
[1+\xi^\c_{12}+\xi^\c_{14}+\xi^\c_{24}+\zeta^\c_{124}]\delta^{\rm D}_{23}
+\hspace{-0.4cm} \prod_{p\in\{1,2,3\}}\hspace{-0.4cm}\{\nbar_p\}
[1+\xi^\c_{12}+\xi^\c_{23}+\xi^\c_{31}+\zeta^\c_{123}]\delta^{\rm D}_{24}
+\hspace{-0.4cm}\prod_{p\in\{1,2,3\}}\hspace{-0.4cm}\{\nbar_p\}
[1+\xi^\c_{12}+\xi^\c_{23}+\xi^\c_{31}+\zeta^\c_{123}]\delta^{\rm D}_{34} \ . \nn\\
& & \label{eq:gamma2}\ea
%
\noindent $\bullet$ {\bf Computing $\langle\Gamma_3\rangle$ : } On
performing the integrations over the delta functions, we have:
\ba
\langle\Gamma_3\rangle & \equiv & \sum_{i_1=i_2\ne i_3=i_4} 
 \hspace{-0.3cm} p(\bx_1,\bx_3, M_1, M_3)\delta^{\rm D}_{12}\delta^{\rm D}_{34} 
+ \hspace{-0.5cm}\sum_{i_1=i_3\ne i_2=i_4}  \hspace{-0.3cm} p(\bx_1,\bx_2, M_1, M_2)
\delta^{\rm D}_{13}\delta^{\rm D}_{24}
+ \hspace{-0.5cm}\sum_{i_1=i_4\ne i_2=i_3}  \hspace{-0.3cm} p(\bx_1,\bx_2, M_1, M_2)
\delta^{\rm D}_{14}\delta^{\rm D}_{23} \nn \\ 
& = &
\frac{}{}\nbar_1\nbar_3\left[1+\xi^\c_{13}\right]\delta^{\rm D}_{12}
\delta^{\rm D}_{34}+\nbar_1\nbar_2\left[1+\xi^\c_{12}\right]
\left[\frac{}{}\delta^{\rm D}_{13}\delta^{\rm D}_{24} +
\delta^{\rm D}_{14}\delta^{\rm D}_{23}\right] \ ,
\label{eq:gamma3}\ea
where to arrive at the second equality, we performed the summations in
the limit of a large number of clusters.

\vspace{0.2cm}
\noindent $\bullet$ {\bf Computing $\langle\Gamma_4\rangle$ : } We
perform the integrations over the delta functions to obtain:
\ba
\langle\Gamma_4\rangle & \hspace{-0.3cm} \equiv \hspace{-0.6cm} & \sum_{i_1=i_2=i_3\ne i_4} 
\hspace{-0.4cm} p(\bx_1,\bx_4, M_1,M_4)\delta^{\rm D}_{12}\delta^{\rm D}_{13}
+ \hspace{-0.5cm} \sum_{i_1=i_2=i_4\ne i_3} \hspace{-0.5cm} p(\bx_1,\bx_3, M_1, M_3)
\delta^{\rm D}_{12}\delta^{\rm D}_{14}
+ \hspace{-0.6cm} \sum_{i_1=i_3=i_4\ne i_2} \hspace{-0.5cm} p(\bx_1,\bx_2, M_1, M_2)
\delta^{\rm D}_{13}\delta^{\rm D}_{14}
+ \hspace{-0.5cm} \sum_{i_2=i_3=i_4\ne i_1} \hspace{-0.5cm} p(\bx_1,\bx_2, M_1, M_2)
\delta^{\rm D}_{23}\delta^{\rm D}_{24} \ .
\nn\ea
Summing and taking the limit of large numbers of clusters we find:
\ba 
\langle\Gamma_4\rangle & = & 
\nbar_1\nbar_4\left[1+\xi^\c_{14}\right]\delta^{\rm D}_{12}\delta^{\rm D}_{13}
+\nbar_1\nbar_3\left[1+\xi^\c_{13}\right]\delta^{\rm D}_{12}\delta^{\rm D}_{14}
+\nbar_1\nbar_2\left[1+\xi^\c_{12}\right]\left\{\delta^{\rm D}_{13}\delta^{\rm D}_{14} +
\delta^{\rm D}_{23}\delta^{\rm D}_{24}\right\} \ .
\label{eq:gamma4}\ea
%
\vspace{0.2cm}
\noindent $\bullet$ {\bf Computing $\langle\Gamma_5\rangle$ : }
Integrating over the delta functions leads to:
\ba
\langle\Gamma_5\rangle & \equiv &  \sum_{i_1=i_2=i_3=i_4} p(\bx_1, M_1)
\delta^{\rm D}_{12}\delta^{\rm D}_{13}\delta^{\rm D}_{14} = 
\frac{}{} \nbar_1 \delta^{\rm D}_{12}\delta^{\rm D}_{13}\delta^{\rm D}_{14} \ .
\label{eq:gamma5}\ea
%
Putting together the terms Eqs.~\ref{eq:gamma1}--\ref{eq:gamma5}
we arrive at the expression:
\ba 
\left<n_{\c, 1}\dots n_{\c, 4}\right> & = & \prod_{p=1}^4 \{\nbar_p\} \left\{\frac{}{}
\hspace{-0.2cm}\left[\frac{}{}\hspace{-0.1cm} 1+\xi^\c_{12}+\xi^\c_{13}+\xi^\c_{14} 
+ \xi^\c_{23}+\xi^\c_{24}+\xi^\c_{34} +\zeta^\c_{123}+\zeta^\c_{234}+\zeta^\c_{341}+\zeta^\c_{412} 
+ \xi^\c_{12}\xi^\c_{34}+\xi^\c_{13}\xi^\c_{24}+\xi^\c_{14}\xi^\c_{23}+\eta^\c_{1234}\right] \right. \nn \\
& &  \left. + \left[\frac{}{}\hspace{-0.1cm} 1+\xi^\c_{13}+\xi^\c_{14}+\xi^\c_{34}
+\zeta^\c_{134}\right]\frac{\delta^{\rm D}_{12}}{\nbar_2}
+ \left[\frac{}{}\hspace{-0.1cm} 1+\xi^\c_{12}+\xi^\c_{24}+\xi^\c_{41}+\zeta^\c_{124}\right]
\frac{\delta^{\rm D}_{23}}{\nbar_3}
+ \left[\frac{}{}\hspace{-0.1cm} 1+\xi^\c_{12}+\xi^\c_{13}+\xi^\c_{23}+\zeta^\c_{123}\right]
\frac{\delta^{\rm D}_{14}}{\nbar_4} \right. \nn \\
& & \left. + \left[\frac{}{}\hspace{-0.1cm} 1+\xi^\c_{12}+\xi^\c_{14}+\xi^\c_{24}
+\zeta^\c_{124}\right] \frac{\delta^{\rm D}_{13}}{\nbar_3}
+ \left[\frac{}{} \hspace{-0.1cm} 1+\xi^\c_{12}+\xi^\c_{23}+\xi^\c_{31}+\zeta^\c_{123}\right]
\frac{\delta^{\rm D}_{r,24}}{\nbar_4}
+ \left[\frac{}{}\hspace{-0.1cm} 1+\xi^\c_{12}+\xi^\c_{23}+\xi^\c_{31}+\zeta^\c_{123}\right]
\frac{\delta^{\rm D}_{34}}{\nbar_4} \right. \nn \\
& & \left. + \left[\frac{}{}\hspace{-0.1cm} 1+\xi^\c_{13}\right]
\frac{\delta^{\rm D}_{12}\delta^{\rm D}_{34}}{\nbar_2\nbar_4}
+ \left[\frac{}{}\hspace{-0.1cm} 1+\xi^\c_{12}\right]
\frac{\delta^{\rm D}_{13}\delta^{\rm D}_{24}}{\nbar_3 \nbar_4}
+ \left[\frac{}{} \hspace{-0.1cm} 1+\xi^\c_{12}\right]
\frac{\delta^{\rm D}_{14}\delta^{\rm D}_{23}}{\nbar_3 \nbar_4}
+ \left[\frac{}{} \hspace{-0.1cm} 1+\xi^\c_{14}\right]
\frac{\delta^{\rm D}_{12}\delta^{\rm D}_{13}}{\nbar_2 \nbar_3}
\right. \nn \\ 
& & \left. + \left[\frac{}{} \hspace{-0.1cm} 1+\xi^\c_{13}\right]
\frac{\delta^{\rm D}_{12}\delta^{\rm D}_{14}}{\nbar_2 \nbar_4}
+ \left[\frac{}{} \hspace{-0.1cm} 1+\xi^\c_{12}\right]
\frac{\delta^{\rm D}_{13}\delta^{\rm D}_{14} + \delta^{\rm D}_{23}\delta^{\rm D}_{24}}
{\nbar_3 \nbar_4}+ \frac{\delta^{\rm D}_{12}\delta^{\rm D}_{13}\delta^{\rm D}_{14}}
{\nbar_2 \nbar_3 \nbar_3} \right\} \ .
\label{eq:L2}
\ea
In the above $\xi^{\c}_{ij}$, $\zeta^{\c}_{ijk}$ and
$\eta^{\c}_{ijkl}$ represent the true three- and four-point connected
correlation functions of galaxy clusters, respectively. Under the
assumption of linear biasing these may be written in terms of the
connected correlation functions of the matter as:
\ba
\xi^\c_{ij}    & \equiv & \xi_{\c}(\bx_i,\bx_j,M_i,M_j) =  b(M_i)b(M_j)\xi_{ij} =  b_ib_j\xi_{ij}\ ;\nn \\
\zeta^\c_{ijk} & \equiv & \zeta_{\c}(\bx_i,\bx_j,\bx_k,M_i,M_j,M_k) = b(M_i)b(M_j)b(M_k)\zeta_{ijk}=
b_ib_jb_k\zeta_{ijk} \ ; \nn \\
\eta^\c_{1234} & \equiv & \eta_{\c}(\bx_1, \bx_2, \bx_3, \bx_4, M_1, M_2, M_3, M_4) 
= b(M_1)b(M_2)b(M_3)b(M_4) \eta_{1234} = b_1 b_2 b_3 b_4 \eta_{1234} \ . \nn 
\ea
Similar to the result of \Eqn{eq:L2}, we can also write down the terms
entering \Eqn{eq:expec1}. Thus we obtain:
\ba 
\left<n_{\c, 1} \,n_{\c, 2} \,n_{\c, 3} \,n_{s, 4}\right> &\equiv&  
\alpha^{-1} \prod_{p=1}^4 \{\nbar_p\} \left\{ \hspace{-0.1cm}\left[\frac{}{}\hspace{-0.1cm}
1+\xi^\c_{12}+\xi^\c_{13}+\xi^\c_{23}+\zeta^\c_{123} \right] \hspace{-0.1cm}
+ \left[\frac{}{}\hspace{-0.1cm} 1+\xi^\c_{13}\right]\hspace{-0.1cm}
\frac{\delta^{\rm D}_{12}}{\nbar_2} 
+ \hspace{-0.1cm}\left[\frac{}{} \hspace{-0.1cm} 1+\xi^\c_{12}\right]\hspace{-0.1cm}
\frac{\delta^{\rm D}_{23} + \delta^{\rm D}_{13}}{\nbar_3} 
+  \frac{\delta^{\rm D}_{12}\delta^{\rm D}_{13}}
{\nbar_2 \nbar_3} \right\} \ ; \nn \\  
\left<n_{\c, 1} \,n_{\c, 2} \,n_{s, 3} \,n_{s, 4}\right> &\equiv &
\alpha^{-2} \prod_{p=1}^4 \{\nbar_p\} \left\{
\left[\frac{}{}\hspace{-0.1cm} 1+\xi^\c_{12} \right]
+ \frac{\delta^{\rm D}_{12}}{\nbar_2}
+ \alpha \left[\frac{}{}1+\xi^\c_{12}\right]\frac{\delta^{\rm D}_{34}}{\nbar_4}
+ \alpha\frac{\delta^{\rm D}_{12}\delta^{\rm D}_{34}}
{\nbar_2 \nbar_4} \right\} \ ; \nn \\
\left<n_{\c, 1} \,n_{s, 2} \,n_{s, 3} \,n_{s, 4}\right> & \equiv & 
\alpha^{-3}  \prod_{p=1}^4 \{\nbar_p\} \left\{ 1 
+ \alpha \left[\frac{\delta^{\rm D}_{23}}{\nbar_3} + \frac{\delta^{\rm D}_{24}}
{\nbar_4} + \frac{\delta^{\rm D}_{34}}{\nbar_4} \right]
+ \alpha^2 \frac{\delta^{\rm D}_{23}\delta^{\rm D}_{24}}
{\nbar_3\nbar_4} \right\} \ ; \nn \\
\left<n_{s, 1} \,n_{s, 2} \,n_{s, 3} \,n_{s, 4}\right> & \equiv & 
\alpha^{-4} \prod_{p=1}^4 \{\nbar_p\} \left\{ 1
+ \alpha \left[\frac{\delta^{\rm D}_{12}}{\nbar_2}
+ \frac{\delta^{\rm D}_{13} + \delta^{\rm D}_{23}}{\nbar_3}
+ \frac{\delta^{\rm D}_{14} + \delta^{\rm D}_{24} + \delta^{\rm D}_{34}}
{\nbar_4} \right]
+ \alpha^{2} \left[ \frac{\delta^{\rm D}_{12}\delta^{\rm D}_{14} + \delta^{\rm D}_{12}\delta^{\rm D}_{34}}
{\nbar_2 \nbar_4}
+ \frac{\delta^{\rm D}_{12}\delta^{\rm D}_{13}}{\nbar_2\nbar_3}
\right. \right.\nn \\
& & + \left.\left.  
\frac{\delta^{\rm D}_{13}\delta^{\rm D}_{24} + \delta^{\rm D}_{14}\delta^{\rm D}_{23}
+ \delta^{\rm D}_{13}\delta^{\rm D}_{14} + \delta^{\rm D}_{23}\delta^{\rm D}_{24}}
{\nbar_3 \nbar_4}
\right]
+\frac{}{}\alpha^3 \frac{\delta^{\rm D}_{12}\delta^{\rm D}_{13}\delta^{\rm D}_{14}}
{\nbar_2 \nbar_3 \nbar_4} \right\} \ .
\ea
Hence, for the second term in \Eqn{eq:expec1} we have:
\ba 
\left<n_{\c, 1} \,n_{\c, 2} \,n_{\c, 3} \,n_{s, 4}\right> +{\rm 3 perms}
& = & \alpha^{-1} \prod_{p=1}^4 \{\nbar_p\}
\left[\frac{}{} \hspace{-0.1cm} 4
+ 2 \left(\frac{}{}\hspace{-0.1cm}\xi^\c_{12} + \xi^\c_{13} + \xi^\c_{23} + \xi^\c_{14} 
+ \xi^\c_{24} + \xi^\c_{34}\right) + \zeta^\c_{123} + \zeta^\c_{124} + \zeta^\c_{134} 
+ \zeta^\c_{234}\right] \nn \\
& & 
+ \left[\frac{}{}\hspace{-0.1cm} 1+\xi^\c_{13}\right]\frac{\delta^{\rm D}_{12}}{\nbar_2}
+ \left[\frac{}{}\hspace{-0.1cm} 1+\xi^\c_{12}\right]\frac{\delta^{\rm D}_{23}}{\nbar_3} 
+ \left[\frac{}{}\hspace{-0.1cm} 1+\xi^\c_{12}\right]\frac{\delta^{\rm D}_{31}}{\nbar_3}
+ \left[\frac{}{}1\hspace{-0.1cm} +\xi^\c_{14}\right]\frac{\delta^{\rm D}_{12}}{\nbar_2}
+ \left[\frac{}{}\hspace{-0.1cm} 1+\xi^\c_{12}\right]\frac{\delta^{\rm D}_{24}}{\nbar_4}
\nn \\ & & 
+ \left[\frac{}{}\hspace{-0.1cm} 1+\xi^\c_{12}\right]\frac{\delta^{\rm D}_{41}}{\nbar_4}
+ \left[\frac{}{}\hspace{-0.1cm} 1+\xi^\c_{14}\right]\frac{\delta^{\rm D}_{13}}{\nbar_3}
+ \left[\frac{}{}\hspace{-0.1cm} 1+\xi^\c_{13}\right]\frac{\delta^{\rm D}_{14}}{\nbar_4}
+ \left[\frac{}{}\hspace{-0.1cm} 1+\xi^\c_{13}\right]\frac{\delta^{\rm D}_{34}}{\nbar_4}
+ \left[\frac{}{}\hspace{-0.1cm} 1+\xi^\c_{24}\right]\frac{\delta^{\rm D}_{23}}{\nbar_3}
\nn \\ & & 
+ \left[\frac{}{}\hspace{-0.1cm} 1+\xi^\c_{23}\right]\frac{\delta^{\rm D}_{24}}{\nbar_4}
+ \left[\frac{}{}\hspace{-0.1cm} 1+\xi^\c_{23}\right]\frac{\delta^{\rm D}_{34}}{\nbar_4}
+ \frac{\delta^{\rm D}_{12}\delta^{\rm D}_{13}}{\nbar_2\nbar_3}
+ \frac{\delta^{\rm D}_{12}\delta^{\rm D}_{14}}{\nbar_2\nbar_4}
+ \frac{\delta^{\rm D}_{13}\delta^{\rm D}_{14}+\delta^{\rm D}_{23}\delta^{\rm D}_{24}}
{\nbar_3\nbar_4} \ .
\ea
For the third term in \Eqn{eq:expec1} we obtain:
\ba 
\left<n_{\c, 1} \,n_{\c, 2} \,n_{s, 3} \,n_{s, 4}\right> + {\rm 5\ perms}
& = & \alpha^{-2} \prod_{p=1}^4 \{\nbar_p\} \left\{
\left[\frac{}{}6+\xi^\c_{12}+\xi^\c_{13}+\xi^\c_{14}+\xi^\c_{23}+\xi^\c_{24}+\xi^\c_{34}\right]
+ \frac{\delta^{\rm D}_{12}}{\nbar_2} + \frac{\delta^{\rm D}_{13} + \delta^{\rm D}_{23}}
{\nbar_3}
+\frac{\delta^{\rm D}_{14}+\delta^{\rm D}_{24}+\delta^{\rm D}_{34}}{\nbar_4} \right.
\nn \\ & &  \left.
+ \alpha \left\{\left[\frac{}{}\hspace{-0.1cm} 1+\xi^\c_{12}\right]\frac{\delta^{\rm D}_{34}}
{\nbar_4} +\left[\frac{}{}\hspace{-0.1cm} 1+\xi^\c_{13}\right]
\frac{\delta^{\rm D}_{24}}{\nbar_4} + \left[\frac{}{}\hspace{-0.1cm} 
1+\xi^\c_{23}\right]\frac{\delta^{\rm D}_{14}}{\nbar_4}
+ \left[\frac{}{}\hspace{-0.1cm} 1+\xi^\c_{14}\right]\frac{\delta^{\rm D}_{23}}{\nbar_3}
+ \left[\frac{}{}\hspace{-0.1cm} 1+\xi^\c_{24}\right]\frac{\delta^{\rm D}_{13}}{\nbar_3}
\right. \right. \nn \\  & &  \left.\left.
+ \left[\frac{}{}\hspace{-0.1cm} 1+\xi^\c_{34}\right]\frac{\delta^{\rm D}_{12}}{\nbar_2} 
+ 2 \frac{\delta^{\rm D}_{12}\delta^{\rm D}_{34}}{\nbar_2\nbar_4} 
+ 2 \frac{\delta^{\rm D}_{13}\delta^{\rm D}_{24} + \delta^{\rm D}_{14}\delta^{\rm D}_{23}}
{\nbar_3 \nbar_4} \right\} \right\}\ .
\ea
For the fourth term in \Eqn{eq:expec1}, we write:
\ba 
\left<n_{\c, 1} \,n_{s, 2} \,n_{s, 3} \,n_{s, 4}\right> + {\rm 3\ perms} 
& \hspace{-0.5cm} = \hspace{-0.4cm} & \alpha^{-3} \prod_{p=1}^4 \{\nbar_p\} \left\{ 4 + 
2 \alpha \hspace{-0.1cm}\left[\frac{\delta^{\rm D}_{12}}{\nbar_2} \hspace{-0.1cm}
+ \frac{\delta^{\rm D}_{13}\hspace{-0.1cm} + \hspace{-0.1cm}\delta^{\rm D}_{23}}{\nbar_3}
+ \frac{\delta^{\rm D}_{14}\hspace{-0.1cm} + \hspace{-0.1cm}\delta^{\rm D}_{24} + \delta^{\rm D}_{34}}
{\nbar_4}\right] +\alpha^2\hspace{-0.1cm} \left[
\frac{\delta^{\rm D}_{13}\delta^{\rm D}_{14} \hspace{-0.1cm}+ \hspace{-0.1cm}
\delta^{\rm D}_{23}\delta^{\rm D}_{24}}{\nbar_3\nbar_4} 
\hspace{-0.1cm}+ \hspace{-0.1cm}\frac{\delta^{\rm D}_{12}\delta^{\rm D}_{14}}{\nbar_2\nbar_4}
\hspace{-0.1cm}+\hspace{-0.1cm} \frac{\delta^{\rm D}_{12}\delta^{\rm D}_{13}}{\nbar_2\nbar_3}
\right] \right\}\ . \nn\\
\ea
On collecting all of the terms and after a little algebra 
we find that \Eqn{eq:FFFF} can be written:
\ba
\left<\Fc(\bx_1)\dots \Fc(\bx_4)\right> & = &
\frac{1}{A^2} \prod_{p=1}^4 \left(\int dM_p w_p\nbar_p\Theta_p\right)
 \left\{ \frac{}{} \hspace{-0.1cm} \eta^\c_{1234} 
+ \zeta^\c_{134}\frac{\delta^{\rm D}_{12}}{\nbar_2}
+ \zeta^\c_{124}\frac{\delta^{\rm D}_{23}}{\nbar_3} 
+ \zeta^\c_{123}\frac{\delta^{\rm D}_{14}}{\nbar_4} 
+ \zeta^\c_{124}\frac{\delta^{\rm D}_{13}}{\nbar_3}
+ \zeta^\c_{123}\frac{\delta^{\rm D}_{24}}{\nbar_4}
\right. \nn \\ & & \left. \hspace{-2.5cm}
+ \zeta^\c_{123}\frac{\delta^{\rm D}_{34}}{\nbar_4}
+ \left[\xi^\c_{12}+\frac{(1+\alpha)}{\nbar_2}\delta^{\rm D}_{12}\right]
 \left[\xi^\c_{34}+\frac{(1+\alpha)}{\nbar_4}\delta^{\rm D}_{34}\right] 
+ \left[\xi^\c_{13}+\frac{(1+\alpha)}{\nbar_3}\delta^{\rm D}_{13}\right]
 \left[\xi^\c_{24}+\frac{(1+\alpha)}{\nbar_4}\delta^{\rm D}_{24}\right]
+ \left[\xi^\c_{14}+\frac{(1+\alpha)}{\nbar_4}\delta^{\rm D}_{14}\right]
\right.\nn \\ & & \left. \hspace{-2.5cm}
\times \left[\xi^\c_{23}+\frac{(1+\alpha)}{\nbar_3}\delta^{\rm D}_{23}\right]
+ \xi^\c_{12}\frac{(\delta^{\rm D}_{13}+\delta^{\rm D}_{23})(\delta^{\rm D}_{14} 
+\delta^{\rm D}_{24})}{\nbar_3\nbar_4} 
+ \xi^\c_{13}\frac{\delta^{\rm D}_{12}(\delta^{\rm D}_{14} +\delta^{\rm D}_{34})}
{\nbar_2\nbar_4}
+ \xi^\c_{14}\frac{\delta^{\rm D}_{12}\delta^{\rm D}_{13}}{\nbar_2\nbar_3}
+ \frac{}{}\hspace{-0.1cm}\frac{(1+\alpha^3) \delta^{\rm D}_{12}\delta^{\rm D}_{13}\delta^{\rm D}_{14}}
{\nbar_2\nbar_3\nbar_4}\right\} \ .
\label{eq:part1}
\ea
The last factor on the right-hand-side of \Eqn{eq:one} can be written:
\ba 
\left<\Fc(\bx_1)\Fc(\bx_3)\right> \left<\Fc(\bx_2)\Fc(\bx_4)\right> 
& = & \prod_{p=1}^4 \left(\int dM_p w_p \nbar_p\Theta_p\right)
\left[\frac{}{}\hspace{-0.1cm}\xi^\c_{13}+\frac{(1+\alpha)\delta^{\rm D}_{13}}{\nbar_3}\right]
\left[\frac{}{}\hspace{-0.1cm}\xi^\c_{24}+\frac{(1+\alpha)\delta^{\rm D}_{24}}{\nbar_4}\right]  
\ . \label{eq:part2}
\ea
Subtracting \Eqn{eq:part2} from \Eqn{eq:part1}, the argument of the
bracket on the right-hand-side of \Eqn{eq:one} can be expressed as:
\ba
\left<\Fc(\bx_1)\dots \Fc(\bx_4)\right> 
- \left<\Fc(\bx_1)\Fc(\bx_3)\right> \left<\Fc(\bx_2)\Fc(\bx_4)\right> & = &
\frac{1}{A^2}\prod_{p=1}^4 \left(\int dM_p w_p \nbar_p \Theta_p\right)
\left\{
\frac{}{}\hspace{-0.1cm}\eta^\c_{1234} + \zeta^\c_{134}\frac{\delta^{\rm D}_{12}}{\nbar_2}
+ \zeta^\c_{124}\frac{\delta^{\rm D}_{23}}{\nbar_3} 
+ \zeta^\c_{123}\frac{\delta^{\rm D}_{14}}{\nbar_4}
\right. \nn \\ & & \hspace{-7cm} \left.
+ \zeta^\c_{124}\frac{\delta^{\rm D}_{13}}{\nbar_3}
+ \zeta^\c_{123}\frac{\delta^{\rm D}_{24}}{\nbar_4}
+ \zeta^\c_{123}\frac{\delta^{\rm D}_{34}}{\nbar_4}
+ \left[\xi^\c_{12}+\frac{(1+\alpha)}{\nbar_2}
\delta^{\rm D}_{12}\right]\left[\xi^\c_{34}+\frac{(1+\alpha)}{\nbar_4}\delta^{\rm D}_{34}\right]
+\left[\xi^\c_{14}+\frac{(1+\alpha)}{\nbar_4}\delta^{\rm D}_{14}\right]
 \left[\xi^\c_{23}+\frac{(1+\alpha)}{\nbar_3}\delta^{\rm D}_{23}\right]
\right. \nn \\ & & \hspace{-7cm} \left.
+ \xi^\c_{12}\frac{(\delta^{\rm D}_{13}+\delta^{\rm D}_{23})(\delta^{\rm D}_{14}
+\delta^{\rm D}_{24})}{\nbar_3\nbar_4}
+ \xi^\c_{13}\frac{\delta^{\rm D}_{12}(\delta^{\rm D}_{14}+\delta^{\rm D}_{34})}
{\nbar_2\nbar_4}
+ \xi^\c_{14}\frac{\delta^{\rm D}_{12}\delta^{\rm D}_{13}}{\nbar_2\nbar_3}
+ \frac{}{}\hspace{-0.1cm} \frac{(1+\alpha^3) \delta^{\rm D}_{12}\delta^{\rm D}_{13}\delta^{\rm D}_{14}}
{\nbar_2\nbar_3\nbar_4}\right\} \ .
\label{eq:CovF0}
\ea
%
\subsection{General expression for the covariance matrix of the power spectrum of $\Fc$}
\label{app:AII.3}
Consider \Eqns{eq:CovF1}{eq:CovF0}, we may Fourier transform all of the
space-dependent terms. For the case of the $n$-point correlation
functions, these are Fourier dual with the $n$-point multi-spectra:
\ba 
\xi_{12} & \equiv & \int \frac{\dq_1}{(2\pi)^3}\frac{\dq_2}{(2\pi)^3}
(2\pi)^3\delta^{\rm D}(\bq_1+\bq_2) P(\bq_1,\bq_2) 
\exp\left[-i\bq_1\cdot\bx_1-i\bq_2\cdot\bx_2 \right]\nn \ ;\\
\zeta_{123} & \equiv & \int \frac{\dq_1}{(2\pi)^3}\frac{\dq_2}{(2\pi)^3}\frac{\dq_3}{(2\pi)^3}
(2\pi)^3\delta^{\rm D}(\bq_1+\bq_2+\bq_3) B(\bq_1,\bq_2,\bq_3)
\exp\left[-i\bq_1\cdot\bx_1-i\bq_2\cdot\bx_2-i\bq_3\cdot\bx_3\right] \ ;\nn \\
\eta_{1234} & \equiv & \int \frac{\dq_1}{(2\pi)^3}\dots \frac{\dq_4}{(2\pi)^3}
(2\pi)^3\delta^{\rm D}(\bq_1+\dots+\bq_4) T(\bq_1,\bq_2,\bq_3,\bq_4) 
\exp\left[-i\bq_1\cdot\bx_1\dots-i\bq_4\cdot\bx_4\right] \ ,
\label{eq:def_high}
\ea
where $B$ and $T$ are bispectrum and trispectrum, respectively. Note
that, owing to the Dirac delta function in the above expressions the
bispectrum and trispectrum are in fact functions of two and three
$k$-vectors, respectively. Using these relations in
\Eqns{eq:CovF1}{eq:CovF0}, we find that the covariance matrix may be
written in general as:
\ba
{\rm Cov}\!\left[|\Fc(\bk_1)|^2,|\Fc(\bk_2)|^2\right] 
& = &
\hspace{-0.2cm}\int \hspace{-0.1cm}\prod_{p=1}^{3}\left(\frac{\dq_p}{(2\pi)^3}\right)
T(\bq_1,\bq_2,\bq_3) \Gf_{(1,1)}(\bk_1-\bq_1)\Gf_{(1,1)}(\bk_2-\bq_2)
\Gf_{(1,1)}(-\bk_1-\bq_3) \Gf_{(1,1)}(-\bk_2+\bq_1+\bq_2+\bq_3) 
\nn \\ & &
+\left|\int \frac{\dq_1}{(2\pi)^3}P(\bq_1)\Gf_{(1,1)}(\bk_1-\bq_1)\Gf_{(1,1)}(\bk_2+\bq_1)
+(1+\alpha)\Gf_{(2,0)}(\bk_1+\bk_2)\right|^2
\nn \\ & & 
+\left|\int \frac{\dq_1}{(2\pi)^3}P(\bq_1)\Gf_{(1,1)}(\bk_1-\bq_1)\Gf_{(1,1)}(-\bk_2+\bq_1)
+(1+\alpha)\Gf_{(2,0)}(\bk_1-\bk_2)\right|^2
\nn \\ & & 
+\int \frac{\dq_1}{(2\pi)^3} \frac{\dq_2}{(2\pi)^3} B(\bq_1,\bq_2)
 \Gf_{(2,1)}(\bk_1+\bk_2-\bq_1)\Gf_{(1,1)}(-\bk_1-\bq_2)\Gf_{(1,1)}(-\bk_2+\bq_1+\bq_2)
\nn \\ & & 
+\int \frac{\dq_1}{(2\pi)^3} \frac{\dq_2}{(2\pi)^3} B(\bq_1,\bq_2)
 \Gf_{(2,1)}(\bk_1-\bk_2-\bq_1)\Gf_{(1,1)}(-\bk_1-\bq_1)\Gf_{(1,1)}(+\bk_2+\bq_1+\bq_2)
\nn \\ & & 
+\int \frac{\dq_1}{(2\pi)^3} \frac{\dq_2}{(2\pi)^3} B(\bq_1,\bq_2)
\Gf_{(2,1)}(-\bk_1+\bk_2-\bq_1)\Gf_{(1,1)}(\bk_1-\bq_1)\Gf_{(1,1)}(-\bk_2+\bq_1+\bq_2) 
\nn \\ & & 
+\int \frac{\dq_1}{(2\pi)^3} \frac{\dq_2}{(2\pi)^3} B(\bq_1,\bq_2)
 \Gf_{(2,1)}(-\bk_2-\bk_1-\bq_1)\Gf_{(1,1)}(\bk_1-\bq_1)\Gf_{(1,1)}(\bk_2+\bq_1+\bq_2)
\nn \\ & & 
+\int \frac{\dq_1}{(2\pi)^3} \frac{\dq_2}{(2\pi)^3} B(\bq_1,\bq_2)
\Gf_{(2,1)}(-\bq_1)\Gf_{(1,1)}(\bk_1-\bq_1)\Gf_{(1,1)}(-\bk_1+\bq_1+\bq_2) 
\nn \\ & & 
+\int \frac{\dq_1}{(2\pi)^3} \frac{\dq_2}{(2\pi)^3} B(\bq_1,\bq_2)
\Gf_{(2,1)}(-\bq_1)\Gf_{(1,1)}(\bk_2-\bq_1)\Gf_{(1,1)}(-\bk_2+\bq_1+\bq_2)
\nn \\ & & 
+\int \frac{\dq_1}{(2\pi)^3} P(\bq_1)\left\{ \left|\Gf_{(2,1)}(\bk_1+\bk_2-\bq_1)\right|^2
+\left|\Gf_{(2,1)}(\bq_1)\right|^2 +\left|\Gf_{(2,1)}(\bk_1-\bk_2-\bq_1)\right|^2\right\}  
\nn \\ & & 
+\int \frac{\dq_1}{(2\pi)^3} P(\bq_1) \left\{\frac{}{}\hspace{-0.1cm}
 \Gf_{(3,1)}(\bk_1-\bq_1)\Gf_{(1,1)}(-\bk_1+\bq_1)
+\Gf_{(3,1)}(\bk_2-\bq_1)\Gf_{(1,1)}(-\bk_2+\bq_1) \, + \right.
\nn \\ & &  \left. 
\frac{}{}\hspace{-0.2cm} \Gf_{(3,1)}(-\bk_1-\bq_1)\Gf_{(1,1)}(\bk_1+\bq_1)
+\Gf_{(3,1)}(-\bk_2-\bq_1)\Gf_{(1,1)}(\bk_2+\bq_1) \right\} \frac{}{}+ (1+\alpha^3)\Gf_{(4,0)}(0) \ .
\label{eq:CovPowerGen}
\ea
\subsection{A result concerning the shell averaging of 
products of the  $\mathcal Q$ functions}
\label{app:AII.4}
Consider the following integral:
\ba 
\int_{V_i} \frac{\dk_1}{V_i} \int_{V_j} \frac{\dk_2}{V_j} 
\tilde\Q_{(j_1, j_2)}^{(i_1, i_2)}(\bk_1+\bk_2) 
\tilde\Q_{(m_1, m_2)}^{(l_1, l_2)}(-\bk_1-\bk_2) 
\hspace{-0.2cm} & = & \hspace{-0.2cm}\int_{V_{i}} 
\frac{\dk_1}{V_i} \int_{V_{j}} \frac{\dk_2}{V_j} \int \dx_1 \,\dx_2 \, 
\mathcal{Q}_{(j_1 \dots j_n)}^{(i_1\dots i_n)}(\bx_1)
\mathcal{Q}_{(m_1 \dots m_n)}^{(l_1\dots l_n)}(\bx_2)
{\rm e}^{i(\bk_1+\bk_2)\cdot(\bx_1-\bx_2)} \nn \\ 
\hspace{-0.2cm}  & = & \hspace{-0.2cm} \int \dx_1 \,\dx_2 \, 
\overline{j_0}(k_i|\bx_1-\bx_2|) \overline{j_0}(k_j|\bx_1-\bx_2|) 
\Q_{(j_1, j_2)}^{(i_1, i_2)}(\bx_1)
\Q_{(m_1, m_2)}^{(l_1, l_2)}(\bx_2) \nn \\ 
\hspace{-0.2cm}  & = & \hspace{-0.2cm} \int \dx_{21} \,
\overline{j_0}(k_ix_{21}) \overline{j_0}(k_jx_{21}) 
\,\Xi_{(j_1, j_2|m_1, m_2)}^{(i_1, i_2|l_1, l_2)}(x_{21}) \ .
\ea
To get the second line in the above equation, we have defined the
shell-averaged spherical Bessel function as
\be \overline{j_0}(k_ix)\equiv \frac{1}{V_i}\int_{k_i-\Delta k/2}^{k_i+\Delta k/2} 
dk k^2 4\pi j_0(k x) \ ,
\nn \ee
and then integrated the exponential functions over the angles
$\widehat{\bk_1}$ and $\widehat{\bk_2}$. The third line in the above
equation resulted from the change of variables $\bx_{21}=\bx_2-\bx_1$,
and from defining the correlation function of the weighted survey
window function as:
\be 
\Xi_{(j_1, j_2|m_1, m_2)}^{(i_1, i_2|l_1, l_2)}(x_{21})
\equiv \int \frac{{\rm d}^2\hat{\bx}_{21}}{4\pi}\int \dx\,
\Q_{(j_1, j_2)}^{(i_1, i_2)}(\bx)
\Q_{(m_1, m_2)}^{(l_1, l_2)}(\bx_{21}+\bx) \ . 
\label{eq:Xidef}
\ee
In the limit that the survey volume is large, the weighted survey
window correlation function is very slowly varying over nearly all
length scales of interest, and so can be approximated by its value at
zero-lag. Hence, we write
\ba
\int_{V_i} \frac{\dk_1}{V_i} \int_{V_j} \frac{\dk_2}{V_j} 
\tilde\Q_{(j_1, j_2)}^{(i_1, i_2)}(\bk_1+\bk_2) 
\tilde\Q_{(m_1, m_2)}^{(l_1, l_2)}(-\bk_1-\bk_2)
\hspace{-0.2cm} & \approx & \hspace{-0.2cm}
\Xi_{(j_1, j_2|m_1, m_2)}^{(i_1, i_2|l_1, l_2)}(0)\int_{V_i} \frac{\dk_1}{V_i}
\int_{V_j} \frac{\dk_2}{V_j} 
 \int_0^{\infty} dx_{21} \,4\pi x^2_{21}j_0(x_{21}k_1)j_0(x_{21}k_2) \nn \\
\hspace{-0.2cm} & = & \hspace{-0.2cm} 
\frac{(2\pi)^3}{V_i}\Xi_{(j_1, j_2|m_1, m_2)}^{(i_1, i_2|l_1, l_2)}(0)
\delta^{K}_{i,j} \ ,
\label{eq:result}
\ea
where for the second line we used the ortogonality relation of the
spherical Bessel functions 
\be
\int_0^{\infty} dx x^2
j_\alpha(ux)j_\alpha(vx) = \frac{\pi}{2u^2}\delta^{\rm D}(u-v) . 
\nn \ee
To evaluate \Eqn{eq:CovGauss4} we need the following expressions:
\ba 
\Xi_{(1,1|1,1)}^{(1,1|1,1)}(0) & = & \int \dx \left[\Q_{(1,1)}^{(1,1)}(\bx)\right]^2 
= \int \dx \left[\G_{(1,1)}(\bx)\right]^4 \ ; \nn \\
\Xi_{(1,1|0)}^{(1,1|2)}(0) & = & \int \dx \Q_{(1,1)}^{(1,1)}(\bx) \Q_{(0)}^{(2)}(\bx)
= \int \dx \left[\G_{(1,1)}(\bx)\right]^2 \G_{(2,0)}(\bx) \ ; \nn \\
\Xi_{(0|0)}^{(2|2)}(0) & = & \int\dx \left[\Q_{(0)}^{(2)}(\bx)\right]^2
= \int \dx \left[\G_{(2,0)}(\bx)\right]^2 \ .
\label{eq:Xiused}
\ea
%
\section{Functional derivatives}\label{app:A3}
To compute $\delta\mathcal{N}[w]$ and $\delta\mathcal{D}[w]$ entering
in the minimization equation \Eqn{eq:minimization}, we first write
down the functional derivatives of the relevant $\Gb$ functions. For a
small variation in the path of $w$, we have:
\ba
\Gb_{(1,1)}[w+\delta w] & = & \int dM \nbar(M) b(M) \Theta(\bx | M)
\left[w(\bx, M) + \delta w(\bx, M)\right]= \Gb_{(1,1)}[w] + \delta\Gb_{(1,1)}[w] \ ; \nn \\
\delta\Gb_{(1,1)}[w] & \equiv & \int dM \nbar(M) b(M) \Theta(\bx | M) \delta w(\bx, M) \ ;
\label{eq:fd_11} \\
\Gb_{(2,0)}[w+\delta w] & = & \int dM \nbar(M) \Theta(\bx | M) \left[w + \delta w\right]^2
= \Gb_{(2, 0)}[w] + \delta \Gb_{(2, 0)}[w] \ ; \nn \\
\delta \Gb_{(2, 0)}[w] & \equiv &  2 \int dM \nbar(M) \Theta(\bx | M) w(\bx, M) \delta w(\bx, M)
\label{eq:fd_20} \ .
\ea
Note that the above calculation is linear in $\delta w$, i.e. we
neglect all terms containing powers higher than $1$ in $\delta w$.  We
also need the functional derivative of the normalization constant $A$
defined by \Eqn{eq:Norm}:
\ba
A[w+\delta w] & = & \int \dx \left\{ \Gb_{(1,1)}[w+\delta w] \right\}^2 = 
\int \dx \left\{ \Gb_{(1,1)}[w] + \delta \Gb_{(1,1)}[w]\right\}^2
= A[w] + 2 \int\,\dx \Gb_{(1,1)}(\bx) \delta\Gb_{(1,1)}[w] = A[w] + \delta A[w]\ ; \nn \\
\delta A[w] & \equiv & 2 \int \dx \,\Gb_{(1,1)}(\bx) \int dM \nbar(M) b(M) 
\Theta(\bx | M) \delta w(\bx, M) \ .
\label{eq:fd_A}
\ea
We now proceed to evaluation the functional derivative of
$\mathcal{N}$ from \Eqn{eq:def_N}:
\be
\delta \mathcal{N}[w] = 2 \int \dx \left[\left[\,\Gb_{(1,1)}(\bx)\right]^2 + 
c\, \Gb_{(2,0)}(\bx)\right] \left[\frac{}{}\hspace{-0.1cm} 2\, \Gb_{(1,1)}(\bx) 
\,\delta \Gb_{(1,1)}[w] + c\, \delta \Gb_{(2,0)}[w] \frac{}{}\hspace{-0.1cm}\right] \ .
\nn \ee
Using the results of \Eqns{eq:fd_11}{eq:fd_20}, we obtain:
\be
\delta \mathcal{N}[w] = 4 \int \dx \,dM \left\{\left[\left[\,\Gb_{(1,1)}(\bx)\right]^2 + 
c\, \Gb_{(2,0)}(\bx)\right] \nbar(M) \Theta(\bx | M) \left[\frac{}{}\hspace{-0.1cm}
\Gb_{(1,1)}(\bx)\,b(M) + c\, w(\bx, M)\frac{}{}\hspace{-0.1cm}\right]\right\}
\delta w(\bx, M) \ .
\label{eq:fd_N}
\ee
Looking at \Eqns{eq:def_D}{eq:fd_A}, we further write down:
\be
\delta \mathcal{D}[w] = 4 A[w] \int \dx\,dM \left\{\Gb_{(1,1)}(\bx) 
\nbar(M) b(M) \Theta(\bx | M)\right\} \delta w(\bx, M) \ .
\label{eq:fd_D}
\ee
%
\section{Proof that $\bbar^2\le \bbarsq$}\label{app:ineq1}
To begin, the Cauchy-Schwarz inequality for two functions $f$ and $g$
states that:
\be \left|\left<f|g\right>\right|^2 \le \left<f|f\right> \left<g|g\right> \ ,\ee
where in the above we are using the following notation:
\be \left<f|g\right> \equiv \int dx q(x) f(x)g(x) \ ,\ee 
and with $q(x)$ an arbitrary positive definite weighting
function. Thus, if we take $f=b(M)$, $g=1$ and $q=\nbar(M)$, then we have
the following inequality:
\be \left[\int_{\Mlim}^{\infty} dM \nbar(M) b(M) \right]^2 
\le 
\left[\int_{\Mlim}^{\infty} dM \nbar(M) b^2(M) \right] 
\left[\int_{\Mlim}^{\infty} dM \nbar(M) \right] \ .\ee 
On dividing both sides of this inequality through by $\nbarh^2$ we
arrive at the stated result.


\end{document}